\documentclass[preprint,showpacs,preprintnumbers,amsmath,amssymb]{revtex4}

\usepackage{graphicx}
\usepackage{dcolumn}
\usepackage{bm}
\usepackage[latin1]{inputenc}

\usepackage{color}

\newcommand{\ud}[1]{{#1^{\dagger}}}
\newcommand{\bra}[1]{\left\langle #1\right|}
\newcommand{\ket}[1]{\left| #1\right\rangle}

\newcommand\Tr{\mathrm{Tr}}

\begin{document}


\title{Luminescence Spectra of Quantum Dots in Microcavities. II.~Fermions.}


\author{Elena del Valle} \affiliation{%
  Departamento de F\'{\i}sica Te\'orica de la Materia Condensada,
  Universidad Aut\'onoma de Madrid, Spain, elena.delvalle@uam.es}%
\author{Fabrice P.~Laussy} \affiliation{%
  School of Physics and Astronomy, University of Southampton,
  Southampton, United Kingdom}%
\author{Carlos Tejedor} \affiliation{%
  Departamento de F\'{\i}sica Te\'orica de la Materia Condensada,
  Universidad Aut\'onoma de Madrid, Spain}%

\date{\today}

\begin{abstract}
  We discuss the luminescence spectra of coupled light-matter systems
  realized with semiconductor heterostructures in microcavities in the
  presence of a continuous, incoherent pumping, when the matter field
  is Fermionic. The linear regime---which has been the main topic of
  investigation both experimentally and theoretically---converges to
  the case of coupling to a Bosonic material field, and has been amply
  discussed in the first part of this work.  We address here the
  nonlinear regime, and argue that, counter to intuition, it is better
  observed at low pumping intensities. We support our discussion with
  particular cases representative of, and beyond, the experimental
  state of the art.  We explore the transition from the quantum to the
  classical regime, by decomposing the total spectrum into individual
  transitions between the dressed states of the light-matter coupling
  Hamiltonian, reducing the problem to the positions and broadenings
  of all possible transitions. As the system crosses to the classical
  limit, rich multiplet structures mapping the quantized energy levels
  melt and turn to cavity lasing and to an incoherent Mollow triplet
  in the direct exciton emission for very good structure. Less ideal
  figures of merit can still betray the quantum regime, with a proper
  balance of cavity versus electronic pumping.
\end{abstract}

\maketitle


\section{Introduction}

In the first part of this work~\cite{elena_laussy08a}, we have
addressed the coupling of light and matter in the particular case
where the material excitation follows Bose statistics, solving
analytically this so-called \emph{linear model} of two harmonic
oscillators. We emphasized how a proper consideration of the
incoherent pumping scheme was needed to describe the effective quantum
state realized in the system, and how this bore consequences on the
spectral lineshapes, in particular on the ability to resolve a Rabi
doublet when the splitting to broadening ratio is small.

In this second part, we turn to the case where the material excitation
follows Fermi statistics and explore the \emph{Jaynes-Cummings
  model}~\cite{jaynes63a}. These two papers focus particularly on
Quantum Dots (QDs) as the matter part of the system, which elementary
excitation---the \emph{exciton}---consists of an electron being
promoted from the valence to the conduction band. The coupling of this
electron with the vacancy it has created in the valence band (the
``hole''), can be either a fermion or a boson, or possibly an
interpolating case of the two~\cite{laussy06b}. Strong Coupling (SC)
regime requires efficient coupling of the dot with light, which can be
enhanced from the dot point of view by increasing the dipole
moment~$d$ of the exciton (the coupling strength~$g$ is proportional
to~$|\mathbf{d}\cdot\mathbf{E}|$ where $\mathbf{E}$ is the cavity
electric field). Large QDs on the other hand would favor the
overlapping of many excitons, whereas small QDs, by confining
separately the electron and hole wavefunctions, fully exhibit Pauli
blocking~\cite{laussy06b}. The former case was studied in part~I and
we now turn to the latter. We use the same formalism and similar
techniques, what allows for a comparison and a clear understanding of
the specifics of both cases. It is known that in absence of
nonlinearity or saturation of some sort, the quantum case is
equivalent to the classical one~\cite{rudin99a}. In particular, the PL
spectrum exhibits a Rabi doublet at resonance, which can be equally
well accounted for by a purely classical model~\cite{zhu90a}. There is
therefore a strong incentive to evidence nonlinear deviations and
attribute them to quantum
effects~\cite{schneebeli08a,steiner08a,srinivasan07a,press07a,kroner08a}.
In studying the coupling of light and matter, be it with atoms or
semiconductors, spontaneous emission from a given initial state has
been overly privileged as the case of
study~\cite{sanchezmondragon83a}. Even when the emitter was modelled
as a two-level system, this configuration allowed to reduce it to the
linear model by considering a single excitation as the initial
state~\cite{carmichael89a,andreani99a,cui06a,auffeves08a,inoue08a}. Most
of the times that the excitation scheme was considered at the same
level as the rest of the dynamics, this was for coherent
pumping~\cite{mollow69a,savage89a,freedhoff94a,clemens00a,barchielli02a,
  florescu06a,bienert07a}. There has been less considerations for the
luminescence spectra under incoherent
pumping~\cite{loffler97a,clemens04a,perea04a,karlovich07a,karlovich08a},
that is the most adequate to describe semiconductors. In the atomic
literature, L\"offler \emph{et al.}~\cite{loffler97a} have considered
spectral shapes for the one-atom laser at resonance by numerical
integration of the master equation and in this context have obtained
some of the lineshapes of the best system that we study
below. Karlovich \emph{et al.}~\cite{karlovich08a} concentrated on
strong coupling at resonance and low pump. In the wake of Part~I, we
rely on semi-analytical results that put clearly apart the spectral
and dynamical aspects of the problem. We shall discuss how our
approach allows in general to track the transitions between the
different regimes and in particular to identify the quantum to
classical one.

One of the most important current task of the SC physics in
semiconductors in our view is the quantitative description of the
experiment with a theory that can provide statistical estimates to the
data, in particular intervals of confidence for the fitting
parameters.  In this respect, there would be little need for fitting
an experiment that would produce a clear observation of the
Jaynes-Cummings energy levels, which is a strong qualitative
effect. But no such structures have been observed so far and the
deviations to Rabi doublet have been understood as non fundamental
features of the problem~\cite{hennessy07a}. The most likely reason for
this lack of crushing observations of the quantum regime in the PL
lineshapes is that the best systems currently available are still
beyond the range of parameters that allows the quantum features to
neatly dominate.  Instead, they are still at a stage where it is easy
to overlook more feeble indications, as shall be seen in what follows
for less ideal systems that are closer to the experimental situation
of today. Another possible reason is that the models are not suitable
and a QD cannot be described by a simple two-level system. Then more
involved theories should take over, with, e.g., full account for
electron and hole band structures and
correlations~\cite{feldtmann06a,gies07a,schneebeli08a}. However, if a
simpler theory is successful, notwithstanding the interest of its more
elaborate and complete counterpart, it clearly facilitates the
understanding and putting the system to useful applications
(especially in a quantum information processing context). At present,
there is more element of chance left in the research for quantum SC
than is actually necessary.  If a quantitative description of even a
``negative experiment'' (not reporting a triplet or quadruplet) could
be provided, this would help tracking and probably even direct the
progress towards the ultimate goal: a fully understood and controlled
SC in the quantum regime.

The rest of this paper is organized as follow. In
Section~\ref{SunNov16134849GMT2008}, we spell out the model. In
Section~\ref{SunNov16135000GMT2008}, we detail the formalism and
provide the expressions for all---and only those---correlation
functions that enter the problem, making it as computationally
efficient as possible for an exact treatment. We provide a
decomposition of the final spectra in terms of transitions of the
dressed states, which gives a clear physical picture of the
problem. In Section~\ref{TueAug12172114CEST2008}, we give the
analytical expressions for the position and broadening of the
resonances of the system at vanishing pumping.  Weighting these
resonances by the self-consistent dynamics of the system established
by finite pumping and decay, gives the final spectral shape. We
discuss in particular the notion of SC that varies from manifold to
manifold, rather than holding for the entire system as a whole.  In
Section~\ref{TueAug12174728CEST2008}, we consider three particular
points representative of the experimental situation, plus one point
beyond what is currently available. We first discuss their behavior
in terms of population and statistical fluctuations as imposed from
the pumping conditions. In Section~\ref{SunNov16135522GMT2008}, we
give the backbone of the final spectra at nonvanishing
excitations. This is the numerical counterpart of
Section~\ref{TueAug12172114CEST2008}, in the presence of arbitrary
pumping. In Section~\ref{SunNov16135934GMT2008}, we present spectral
shapes for the three points in a variety of configuration and compare
them to each other. In Section~\ref{SunNov16140046GMT2008}, we
investigate the situation at nonzero detuning, which is a case of
particular importance in semiconductor physics. Finally, in
Section~\ref{FriNov7121705GMT2008}, we provide an overview of the
results and conclude.

\section{Model}
\label{SunNov16134849GMT2008}

The Hamiltonian that describes a Fermionic QD in strong coupling with
a single-mode microcavity is the Jaynes-Cummings
Hamiltonian~\cite{jaynes63a} ($\hbar$ is taken as 1 along the paper):
\begin{equation}
  \label{eq:TueJul15133406CEST2008}
  H=\omega_a\ud{a}a+\omega_\sigma\ud{\sigma}\sigma+g(\ud{a}\sigma+a\ud{\sigma})
\end{equation}
with~$a$ the cavity mode annihilation operator tuned at
energy~$\omega_a$ (obeying Bose statistics)
and~$\sigma=(\sigma_x+i\sigma_y)/2$ the exciton annihilation operator
at energy~$\omega_\sigma$ (obeying Fermi statistics, $\sigma_{x,y}$
are Pauli matrices). The two modes are coupled with the interaction
strength~$g$ and close enough to resonance to allow for the rotating
wave approximation~\cite{carmichael_book02a}. The detuning between the
modes is defined as $\Delta=\omega_a-\omega_\sigma$. The Liouvillian
to describe the system in the framework of a quantum dissipative
master equation, $\partial_t\rho=\mathcal{L}\rho$, has the same form
as that in Part~I of this work~\cite{elena_laussy08a}:
\begin{subequations}
  \label{eq:ThuOct18162449UTC2007}
  \begin{align}
    \mathcal{L} O=i[O,H]&+\sum_{c=a,\sigma}\frac{\gamma_c}2(2cO\ud{c}-\ud{c}cO-O\ud{c}c)\\
    &+\sum_{c=a,\sigma}\frac{P_c}2(2\ud{c}Oc-c\ud{c}O-Oc\ud{c})\,,
  \end{align}
\end{subequations}
where~$\rho$ is the density matrix for the combined
Fermi-emitter/cavity system.  The only change in both
Eqs.~(\ref{eq:TueJul15133406CEST2008})
and~(\ref{eq:ThuOct18162449UTC2007}) with respect to their counterpart
in Ref.~\cite{elena_laussy08a} is the replacement of the Boson
operator~$b$ (as it was called in part~I) to describe the matter
field, by a Fermion operator: $b\leftrightarrow\sigma$. This
interchange has far reaching consequences, as will be seen in the
following of this text.

We shall not focus on the difference between the spontaneous emission
(SE) of an initial state in absence of any pumping, and the steady
state (SS) established in presence of this pumping, as we did in part
I for the Boson case.
SS is the most relevant case for the experimental configuration that
we have in mind. Rather than contrasting the SE/SS results, as was
done in Part~I, we shall therefore contrast the Boson/Fermion
cases. For this reason and for concision, we shall not use the ``SS''
superscript and assume that which of the SE/SS case is assumed is
clear from context or from the presence of the time variable~$t$.

In the Boson case, the quantum state of the system is not by itself an
interesting quantity as most of its features are contained in its
reduced density matrices, that are simply and in all cases thermal
states with effective temperatures specified by the mean populations
of the modes~$n_a$ and~$n_\sigma$~\cite{elena_laussy08a}, defined by:
\begin{equation}
  \label{eq:SunJul20214055CEST2008}
  n_a=\langle\ud{a}a\rangle\quad\text{and}\quad n_\sigma=\langle\ud{\sigma}\sigma\rangle\,.
\end{equation}
For this reason, the higher order correlator
\begin{equation}
  \label{eq:SunJul20213709CEST2008}
  g^{(2)}={\langle\ud{a}\ud{a}aa\rangle}/{n_a^2}\,,
\end{equation}
that measures the fluctuations in the photon numbers, does not contain
any new information.  In the Fermion case however, $g^{(2)}$ becomes
nontrivial, because the saturation of the dot provides a nonlinearity
in the system that can produce various types of statistics, from the
coherent Poisson distributions, encountered in lasers (where the
nonlinearity is provided by the feedback and laser gain), to
Fock-state statistics, with antibunching, exhibited by systems with a
quantum state that has no classical counterpart. The fluctuations in
particle numbers influence the spectral shape. The full statistics
itself is most conveniently obtained from the master equation with
elements $\rho_{m,i;n,j}$ for $m$, $n$ photons and~$i$, $j$ exciton
($m, n\in\mathbf{N}$, $i, j\in\{0,1\}$). The distribution function of
the photon number is simply
$\mathrm{p}[n]=\rho_{n,0;n,0}+\rho_{n,1;n,1}$.

Rather than to consider the equations of motion for the matrix
elements directly, it is clearer and more efficient to consider only
elements that are nonzero in the steady state. These are:
\begin{equation}
  \label{eq:SunJul20203845CEST2008}
  \mathrm{p}_0[n]=\rho_{n,0;n,0}\,,\quad \mathrm{p}_1[n]=\rho_{n,1;n,1}\quad\text{and}\quad \mathrm{q}[n]=\rho_{n,0;n-1,1}\,,
\end{equation}
and correspond to, respectively, the probability to have~$n$ photons
with ($\mathrm{p}_1$) or without ($\mathrm{p}_0$) exciton, and the
coherence element between the states~$\ket{n,0}$ and~$\ket{n-1,1}$,
linked by the SC Hamiltonian. Both~$p_0$ and~$p_1$ are real. It is
convenient to separate~$q$ into its real and imaginary parts,
$\mathrm{q}[n]=\mathrm{q}_\mathrm{r}[n]+i \mathrm{q}_\mathrm{i}[n]$ as
they play different roles in the dynamics. The equations for these
quantities, derived from Schr\"odinger equation for the Liouvillian
Eq.~(\ref{eq:ThuOct18162449UTC2007}), read:
\begin{subequations}
  \label{eq:SunJul20204529CEST2008}
  \begin{align}
    \partial_t\mathrm{p}_0[n]=&-\big((\gamma_a+P_a)n+P_a+P_\sigma\big)\mathrm{p}_0[n]+\gamma_a(n+1)\mathrm{p}_0[n+1]+P_an\mathrm{p}_0[n-1]\label{eq:SunNov23185347GMT2008}\nonumber\\
    &+\gamma_\sigma \mathrm{p}_1[n]-2g\sqrt{n}\mathrm{q}_\mathrm{i}[n]\,,\\
    \partial_t\mathrm{p}_1[n]=&-\big((\gamma_a+P_a)n+P_a+\gamma_\sigma\big)\mathrm{p}_1[n]+\gamma_a(n+1)\mathrm{p}_1[n+1]+P_an\mathrm{p}_1[n-1]\nonumber\\
    &+P_\sigma \mathrm{p}_0[n]+2g\sqrt{n+1}\mathrm{q}_\mathrm{i}[n+1]\,,\label{eq:SunNov23185436GMT2008}\\
    \partial_t\mathrm{q}_\mathrm{i}[n]=&-\Big((\gamma_a+P_a)n-\frac{\gamma_a-P_a}{2}+\frac{\gamma_\sigma+P_\sigma}{2}\Big)\mathrm{q}_\mathrm{i}[n]\nonumber\\
    &+\gamma_a\sqrt{(n+1)n}\mathrm{q}_\mathrm{i}[n+1]+P_a\sqrt{(n-1)n}\mathrm{q}_\mathrm{i}[n-1]\nonumber\\
    &+g\sqrt{n}(\mathrm{p}_0[n]-\mathrm{p}_1[n-1])-\Delta
    \,\mathrm{q}_\mathrm{r}[n]\,,\label{eq:SunNov23185449GMT2008}\\
    \partial_t\mathrm{q}_\mathrm{r}[n]=&-\Big((\gamma_a+P_a)n-\frac{\gamma_a-P_a}{2}+\frac{\gamma_\sigma+P_\sigma}{2}\Big)\mathrm{q}_\mathrm{r}[n]\nonumber\\
    &+\gamma_a\sqrt{(n+1)n}\mathrm{q}_\mathrm{r}[n+1]+P_a\sqrt{(n-1)n}\mathrm{q}_\mathrm{r}[n-1]+\Delta \,\mathrm{q}_\mathrm{i}[n]\,.
  \end{align}
\end{subequations}
Note that in the steady state, Eqs.~(\ref{eq:SunJul20204529CEST2008})
are detailed-balance type of equations. The conditional photon
statistics with and without the exciton are similar, and coupled
through the imaginary part of the~$\mathrm{q}$ distribution (that is
not a probability). At resonance, the real part of the coherence
distribution, $\mathrm{q}_\mathrm{r}$, gets decoupled and vanishes in
the steady state. As a result, only
Eqs.~(\ref{eq:SunNov23185347GMT2008})--(\ref{eq:SunNov23185449GMT2008})
need to be solved. When~$g$ vanishes, $\mathrm{q}_\mathrm{i}$ does not
couple the two modes anymore, and their statistics become thermal like
in the boson case. Through the off-diagonal
elements~$\mathrm{q}_\mathrm{i}$, the photon density matrix can vary
between Poissonian, thermal (superpoissonian) and subpoissonian
distributions~\cite{delvalle07b}.

\section{Correlation functions and Spectra}
\label{SunNov16135000GMT2008}

The main quantity of interest of this paper is the luminescence
spectrum of the system. In the Boson case, the
symmetry~$a\leftrightarrow b$ allowed to focus exclusively on the
cavity-emission without loss of generality, as the direct exciton
emission could be obtained from the cavity emission by interchanging
parameters. Here, the exciton (Fermion) and photon (Boson) are
intrinsically different, and no simple relationship links them. They
must therefore be computed independently:
\begin{align}
  \label{eq:MonJul21132824CEST2008}
  S_c(\omega)&=\frac{1}{\pi n_c}\lim_{t\rightarrow\infty}\Re\int_{0}^{\infty}\langle\ud{c}(t)c(t+\tau)\rangle e^{i\omega\tau}d\tau\,,\quad\text{with~$c=a, \sigma$}\,.
\end{align}
We consider normalized spectra for convenient comparisons of the
lineshapes. The normalizing factor is the population~$n_c$, as seen
straightforwardly from $\int S_c(\omega)d\omega=1$. 

As in Part~I, we recourse to the quantum regression theorem to compute
the two times average~$\langle\ud{c}(t)c(t+\tau)\rangle$. We first
identify the set of closing operators~$C_{\{\eta\}}$ in the sense
that, for any operator~$\Omega$, the
relation~$\Tr(C_{\{\eta\}}\mathcal{L}\Omega)=\sum_{\{\lambda\}}
M_{\{\eta\lambda\}}\Tr(C_{\{\lambda\}}\Omega)$ is satisfied for
some~$M_{\{\eta\lambda\}}$ to be specified. In the linear case, the
set of $a^mb^n$ with~$m$, $n\in\mathbf{N}$ is closed, what allows for
an analytical solution. In the nonlinear case, four indices are
required to label the closing operators, namely
$\{\eta\}=(m,n,\mu,\nu)$ in
$C_{\{\eta\}}=\ud{a}^ma^n\ud{\sigma}^\mu\sigma^\nu$ with~$m$,
$n\in\mathbf{N}$ and~$\mu$, $\nu\in\{0,1\}$.  The links established
between them by the Liouvillian dynamics are given by
$\Tr(\ud{a}^ma^n\ud{\sigma}^\mu\sigma^\nu{\mathcal
  L}\Omega)=\sum_{pq\pi\theta}M_{\substack{mn\mu\nu\\pq\pi\theta}}\Tr(\ud{a}^p
a^q\ud{\sigma}^\pi\sigma^\theta\Omega)$, with~$M$ defined as:
\begin{subequations}
  \label{eq:SunOct21120346UTC2007}
  \begin{align}
    &M_{\substack{mn\mu\nu\\mn\mu\nu}}=i\omega_a(m-n)+i\omega_\sigma(\mu-\nu)-\frac{\gamma_a-P_a}2(m+n)-\frac{\gamma_\sigma+P_\sigma}2(\mu+\nu)\,,\label{eq:ThuAug7220550CEST2008}\\
    &M_{\substack{mn\mu\nu\\m-1,n-1,\mu\nu}}={P_a}mn\,,\quad
    M_{\substack{mn\mu\nu\\mn,1-\mu,1-\nu}}={P_\sigma}\mu\nu\,,\\
    &M_{\substack{mn\mu\nu\\m-1,n,1-\mu,\nu}}=M^*_{\substack{nm\nu\mu\\n,m-1,\nu,1-\mu}}=igm(1-\mu)\,,\\
    &M_{\substack{mn\mu\nu\\m,n+1,\mu,1-\nu}}=M^*_{\substack{nm\nu\mu\\n+1,m,1-\nu,\mu}}=-ig\nu\,,\\
    &M_{\substack{mn\mu\nu\\m,n+1,1-\mu,\nu}}=M^*_{\substack{nm\nu\mu\\n+1,m,\nu,1-\mu}}=2ig\nu(1-\mu)\,,
  \end{align}
\end{subequations}
and zero everywhere else.

We are interested in this text in~$\Omega=\ud{c}$ with $c=a$
and $\{\eta_a\}=(0,1,0,0)$ on the one hand, to get the equation
for~$\langle\ud{a}(t)a(t+\tau)\rangle$ that will provide the cavity
emission spectrum, and~$c=\sigma$
with~$\{\eta_\sigma\}=(0,0,0,1)$ on the other hand, to get the
equation for $\langle\ud{\sigma}(t)\sigma(t+\tau)\rangle$ for the QD
direct emission spectrum. Contrary to the Boson case, this procedure
leads to an infinite set of coupled equations. The equations of motion
for both $\langle \ud{a}(t)C_{(0, 1, 0, 0)}(t+\tau)\rangle$
and~$\langle\ud{\sigma}(t)C_{(0,0,0,1)}(t+\tau)\rangle$ involve the
same family of closing operators~$C_{\{\eta\}}$, namely
with~$\eta\in\bigcup_{k\ge1}\mathcal{N}_k$
where~$\mathcal{N}_1=\{(0,1,0,0),(0,0,0,1)\}$ the manifold of the
boson case, and for~$k>1$:
\begin{equation}
  \label{eq:WedAug6172513CEST2008}
  \mathcal{N}_k=\{(k-1,k,0,0),(k-1,k-1,0,1),(k-2,k,1,0),(k-2,k-1,1,1)\}\,.
\end{equation}

\begin{figure}[hbpt]
  \centering
  \includegraphics[width=.75\linewidth]{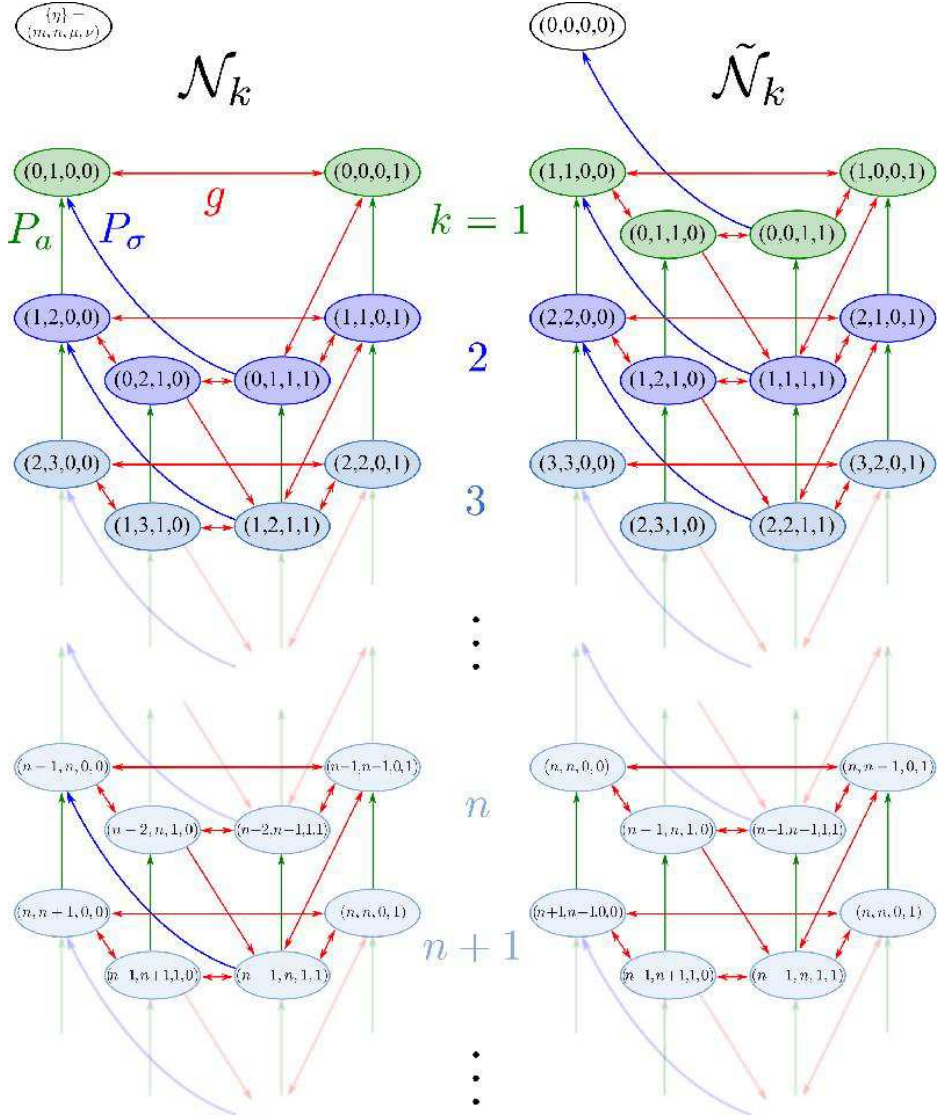}
  \caption{(Color online) Chain of correlators---indexed
    by~$\{\eta\}=(m,n,\mu,\nu)$---linked by the dissipative
    Jaynes-Cummings dynamics. On the left (resp., right), the
    set~$\bigcup_k\mathcal{N}_k$ (resp.,
    $\bigcup_k\tilde{\mathcal{N}}_k$) involved in the equations of the
    two-time (resp., single-time) correlators. In green are shown the
    first manifolds~$\mathcal{N}_1$ and $\tilde{\mathcal{N}}_1$ that
    correspond to the Boson case, and in increasingly lighter shades
    of blues, the higher manifolds~$\mathcal{N}_k$
    and~$\tilde{\mathcal{N}}_k$. The equation of motion~$\langle
    \ud{a}(t)C_{\{\eta\}}(t+\tau)\rangle$ (resp. $\langle
    \ud{\sigma}(t)C_{\{\eta\}}(t+\tau)$) with $\eta\in\mathcal{N}_k$
    requires for its initial value the correlator~$\langle
    C_{\{\tilde\eta\}}\rangle$
    with~$\{\tilde\eta\}\in\tilde{\mathcal{N}}_k$ defined
    from~$\{\eta\}=(m,n,\mu,\nu)$ by~$\{\tilde\eta\}=(m+1,n,\mu,\nu)$
    (resp. $(m,n,\mu+1,\nu)$), as seen on the diagram.  The red arrows
    indicate which elements are linked by the coherent (SC) dynamics,
    through the coupling strenght~$g$, while the green/blue arrows
    show the connections due to the incoherent cavity/exciton
    pumpings, respectively. The self-coupling of each node to itself
    is not shown. This is where~$\omega_{a,\sigma}$
    and~$\gamma_{a,\sigma}$ enter.}
  \label{fig:ThuJul24124951CEST2008}
\end{figure}

The links between the various correlators tracked through the
indices~$\{\eta\}$, are shown in
Fig.~\ref{fig:ThuJul24124951CEST2008}. To solve the differential
equations of motion, the initial value of each correlator is also
required, e.g., $\langle \ud{a}(t)a(t+\tau)\rangle$
demands~$\langle(\ud{a}a)(t)\rangle$, etc. The initial values of
$\langle a(t)C_{\{\eta\}}(t+\tau)\rangle$ (resp., $\langle
\sigma(t)C_{\{\eta\}}(t+\tau)\rangle$) can be conveniently computed
within the same formalism, recurring to~$\Omega=1$
and~$C_{\{\tilde\eta\}}$ with~$\{\tilde\eta\}=\{m+1,n,\mu,\nu\}$
(resp., $\{m,n,\mu+1,\nu\}$).  This allows to compute also the
single-time dynamics $\langle C_{\{\tilde\eta\}}(t)\rangle$, and their
steady state, from the same tools used as for the two-time dynamics
through the quantum regression theorem. The indices $\{\tilde\eta\}$
required for the single-time correlators form a set---that we call
$\tilde{\mathcal{N}}=\bigcup_{k\ge1}\tilde{\mathcal{N}}_k$---that is
disjoint from $\bigcup_{k\ge1}\mathcal{N}_k$, required for the
two-times dynamics.  The set~$\tilde{\mathcal{N}}$ has---beside the
constant term~$\{\eta_0\}=(0,0,0,0)$---two more elements for the lower
manifold (of the Boson case). This is because $\{\eta_a\}=(0,1,0,0)$
and~$\{\eta_\sigma\}=(0,0,0,1)$ invoke~$(1,1,0,0)$ and~$(1,0,0,1)$ for
the cavity spectrum on the one hand, and~$(0,1,1,0)$ and~$(0,0,1,1)$
for the exciton emission on the other. At higher orders~$k>1$, all
two-times correlators~$\mathcal{N}_k$ otherwise depend on the same
four single-time correlators~$\tilde{\mathcal{N}}_k$. Independently of
which spectrum one wishes to compute, these four elements~$(1,1,0,0)$,
$(1,0,0,1)$, $(0,1,1,0)$ and~$(0,0,1,1)$ of~$\tilde{\mathcal{N}}_1$
are needed in all cases as they are linked to each other, as shown on
Fig.~\ref{fig:ThuJul24124951CEST2008}.

On the figure, only the type of coupling---coherent, through~$g$, or
incoherent, through the pumpings~$P_{a,\sigma}$---has been
represented. Weighting coefficients are given by
Eqs.~(\ref{eq:SunOct21120346UTC2007}). Of particular relevance is the
self-coupling of each correlator to itself, not shown on the figure
for clarity. Its coefficient, Eq.~(\ref{eq:ThuAug7220550CEST2008}),
lets enter~$\gamma_{a,\sigma}$ that do not otherwise couple any one
correlator to any of the others. This makes it possible to describe
decay by simply providing an imaginary part to the Energy in
Eq.~(\ref{eq:TueJul15133406CEST2008}). The incoherent pumping, on the
other hand, establishes a new set of connections between
correlators. Note, however, that at the exception of~$\{\eta_0\}$, the
pumping does not enlarge the sets~$\bigcup\mathcal{N}_k$,
$\bigcup\tilde{\mathcal{N}}_k$: the structure remains the same (also,
technically, the computational complexity is identical), only with the
correlators affecting each other differently. The addition
of~$\{\eta_0\}$ by the pumping terms bring the same additional physics
in the Boson and Fermion cases: it imposes a self-consistent steady
state over a freely chosen initial condition.  In the Boson case, the
pumping had otherwise only a direct influence in renormalizing the
self-coupling of each correlator. In the Fermion case, it brings
direct modifications to the Jaynes-Cummings coherent dynamics. But its
contribution to the self-coupling is also important, and gives rise to
an interesting fermionic opposition to the bosonic effects as seen in
Eq.~($\ref{eq:ThuAug7220550CEST2008}$) in the effective linewidth:
\begin{equation}
  \label{eq:MonAug11125316CEST2008}
  \Gamma_{a}=\gamma_{a}-P_a\,,\qquad\Gamma_\sigma=\gamma_\sigma+P_\sigma\,.
\end{equation}
For later convenience, we also define:
\begin{equation}
  \label{eq:TueAug12190422CEST2008}
  \Gamma_\pm=\frac{\Gamma_a\pm\Gamma_b}4\,.
\end{equation}
In Eq.~(\ref{eq:MonAug11125316CEST2008}), it is seen that, whereas the
incoherent cavity pumping narrows the linewidth, as a manifestation of
its boson character, the incoherent exciton pumping broadens it. This
opposite tendencies bear a capital importance for the lineshapes, as
narrow lines favor the observation of a structure, whereas broadening
hinders it. On the other hand, the cavity incoherent pumping always
results in a thermal distribution of photons with large fluctuations
of the particle numbers, that result in an inhomogeneous broadening,
whereas the exciton pumping can grow a Poisson-like distribution with
little fluctuations. Both types of pumping, however, ultimately bring
decoherence to the dynamics and induce the transition into weak
coupling (WC), with the lines composing the spectrum collapsing into
one. Putting all these effects together, there is an optimum
configuration of pumpings where particle fluctuations compensate for
the broadening of the interesting lines, enhancing their resolution in
the spectrum, as we shall see when we discuss the results below.

As there is no finite closure relation, some truncation is in
order. We will adopt the scheme where a maximum of~$n_\mathrm{t}$
excitation(s) (photon plus excitons) is allowed at the
$n_\mathrm{t}$th order, thereby truncating into manifolds of
excitations, which is the most relevant picture. The exact result is
recovered in the limit $n_t\rightarrow\infty$. As seen in
Fig.~\ref{fig:ThuJul24124951CEST2008}, the number~$s_\mathrm{t}$ of
two-time correlators from~$\mathcal{N}$ up to order~$n_\mathrm{t}$ is
$s_\mathrm{t}=4n_\mathrm{t}-2$ and the number of mean values
from~$\tilde{\mathcal{N}}$ is~$4n_\mathrm{t}$. The problem is
therefore computationally linear in the number of excitations, and as
such is as simple as it could be for a quantum system.  The general
case consists in a linear system of~$s_\mathrm{t}$ coupled
differential equations, whose matrix of coefficients [specified by
Eqs.~(\ref{eq:SunOct21120346UTC2007})] is, in the basis
of~$C_{\{\eta\}}$, a~$s_\mathrm{t}\times s_\mathrm{t}$ square matrix
that we denote~$\mathbf{M}$. With these definitions, the quantum
regression theorem becomes:
\begin{equation}
  \label{eq:ThuJul24152824CEST2008}
  \partial_\tau\mathbf{v}_c(t,t+\tau)=\mathbf{M}\mathbf{v}_c(t,t+\tau)  
\end{equation}
where $\mathbf{v}_c(t,t+\tau)=\langle
\ud{c}(t)\mathbf{C}_{\{\eta\}}(t+\tau)\rangle$. Explicitly, for the
lower manifolds, e.g., for~$c=a$:
\begin{equation}
  \label{eq:TueAug12131413CEST2008}
  \mathbf{C}_{\{\eta\}}=
  \begin{pmatrix}
    C_{(0,1,0,0)}\\
    C_{(0,0,0,1)}\\
    C_{(1,2,0,0)}\\
    C_{(1,1,0,1)}\\
    C_{(0,2,1,0)}\\
    \vdots
  \end{pmatrix}
  \quad\text{and}\quad
  \mathbf{v}_{a}(t,t+\tau)=
  \begin{pmatrix}
    \langle \ud{a}(t)a(t+\tau)\rangle\\
    \langle \ud{a}(t)\sigma(t+\tau)\rangle\\
    \langle \ud{a}(t)(\ud{a}a^2)(t+\tau)\rangle\\
    \langle \ud{a}(t)(\ud{a}a\sigma)(t+\tau)\rangle\\
    \langle \ud{a}(t)(a^2\ud{\sigma})(t+\tau)\rangle\\
    \vdots
  \end{pmatrix}\,.
\end{equation}

The ordering of the correlators is arbitrary. We fix it to that of
Fig.~\ref{fig:ThuJul24124951CEST2008}, as seen in
Eq.~(\ref{eq:TueAug12131413CEST2008}). With this convention, the
indices of the two correlators of interests are:
\begin{equation}
  \label{eq:TueAug12133655CEST2008}
  i_a=1,\quad i_\sigma=2\,.
\end{equation}

To solve Eq.~(\ref{eq:ThuJul24152824CEST2008}), we introduce the
matrix~$\mathbf{E}$ of normalized eigenvectors of~$\mathbf{M}$,
and~$-\mathbf{D}$ the diagonal matrix of eigenvalues:
\begin{equation}
  \label{eq:ThuJul24161518CEST2008}
  -\mathbf{D}=\mathbf{E}^{-1}\mathbf{M}\mathbf{E}\,.
\end{equation}

The formal solution is then
$\mathbf{v}_c(t,t+\tau)=\mathbf{E}e^{-\mathbf{D}\tau}\mathbf{E}^{-1}\mathbf{v}_c(t,t)$.
Integration of~$\int e^{(-\mathbf{D}+i\omega)\tau}d\tau$ and
application of the Wiener-Khintchine formula yield for the~$i_a$th and
$i_\sigma$th rows of~$\mathbf{v}_c$ the emission spectra of the
cavity, $S_a=\frac{1}{\pi n_a}\Re\int\langle
\ud{a}(t){a}(t+\tau)\rangle e^{i\omega\tau}\,d\tau$, and of the direct
exciton emission, $S_\sigma=\frac{1}{\pi
  n_\sigma}\Re\int\langle\ud{\sigma}(t){\sigma}(t+\tau)\rangle
e^{i\omega\tau}\,d\tau$, respectively. We find, to
order~$n_\mathrm{t}$:
\begin{equation}
  \label{eq:SatAug2112118CEST2008}
  S_c(\omega)=\frac{1}{\pi}\Re\sum_{p=1}^{s_\mathrm{t}}\frac{L^c_{i_cp}+iK^c_{i_cp}}{D_p-i\omega}\,,\quad c=a,\sigma\,,
\end{equation}
where $L^c_{i_cp}$ and~$K^c_{i_cp}$ are the real and the imaginary
part, respectively, of
$[\mathbf{E}]_{i_cp}[\mathbf{E}^{-1}\mathbf{v}_c(t,t)]_p/n_c$:
\begin{equation}
  \label{eq:SatAug2112438CEST2008}
  L^c_{i_cp}+iK^c_{i_cp}=\frac1{n_c}[\mathbf{E}]_{i_cp}\sum_{q=1}^{s_\mathrm{t}}[\mathbf{E}^{-1}]_{pq}[\mathbf{v}_c(t,t)]_q\,,\quad 1\le p\le s_\mathrm{t}\,,
\end{equation}
and $D_p=[\mathbf{D}]_{pp}$ (when we refer to elements of a matrix or
a vector by its indices, we enclose it with square brackets to
distinguish from labelling indices). Further defining $\gamma_p$
and~$\omega_p$ as the real and imaginary parts, respectively, of~$D_p$
\begin{equation}
  \label{eq:SatAug2111206CEST2008}
  \gamma_p+i\omega_p=D_p\,,
\end{equation}
we can write Eq.~(\ref{eq:SatAug2112118CEST2008}) in a less concise
but more transparent way. To all orders, it reads:
\begin{equation}
\label{eq:FriOct26050345UTC2007}
S_c(\omega)=\frac1{\pi}\lim_{n_\mathrm{t}\rightarrow\infty}\sum_{p=1}^{s_\mathrm{t}}\left(L^c_{i_cp}\frac{\gamma_p}{(\omega-\omega_p)^2+\gamma_p^2}-K^c_{i_cp}\frac{\omega-\omega_p}{(\omega-\omega_p)^2+\gamma_p^2}\right)\,.
\end{equation}

Equation~(\ref{eq:FriOct26050345UTC2007}) brings together all the
important quantities that define the luminescence spectrum of a
quantum dot in a microcavity. The lineshape is composed of a series of
Lorentzian and Dispersive parts, whose positions and broadenings are
specified by $\omega_p$ and~$2\gamma_p$,
cf.~Eq.~(\ref{eq:SatAug2111206CEST2008}), and which are weighted by
the coefficients~$L^c_{i_cp}$ and~$K^c_{i_cp}$,
cf.~Eq.~(\ref{eq:SatAug2112438CEST2008}).  The former pertain to the
structure of the spectral shape as inherited from the Jaynes-Cummings
energy levels. They are, as such, independent of the channel of
detection (cavity or direct exciton emission).  We devote
Section~\ref{TueAug12172114CEST2008} to them. The latter reflect the
quantum state that has been realized in the system under the interplay
of pumping and decay. They determine which lines actually appear in
the spectra, and with which intensity. Naturally, the channel of
emission is a crucial element in this case. We devote
Section~\ref{TueAug12174728CEST2008} to this aspect of the problem.

\section{Spectral structure}
\label{TueAug12172114CEST2008}

\begin{figure}[htbp]
  \centering
  \includegraphics[width=.9\linewidth]{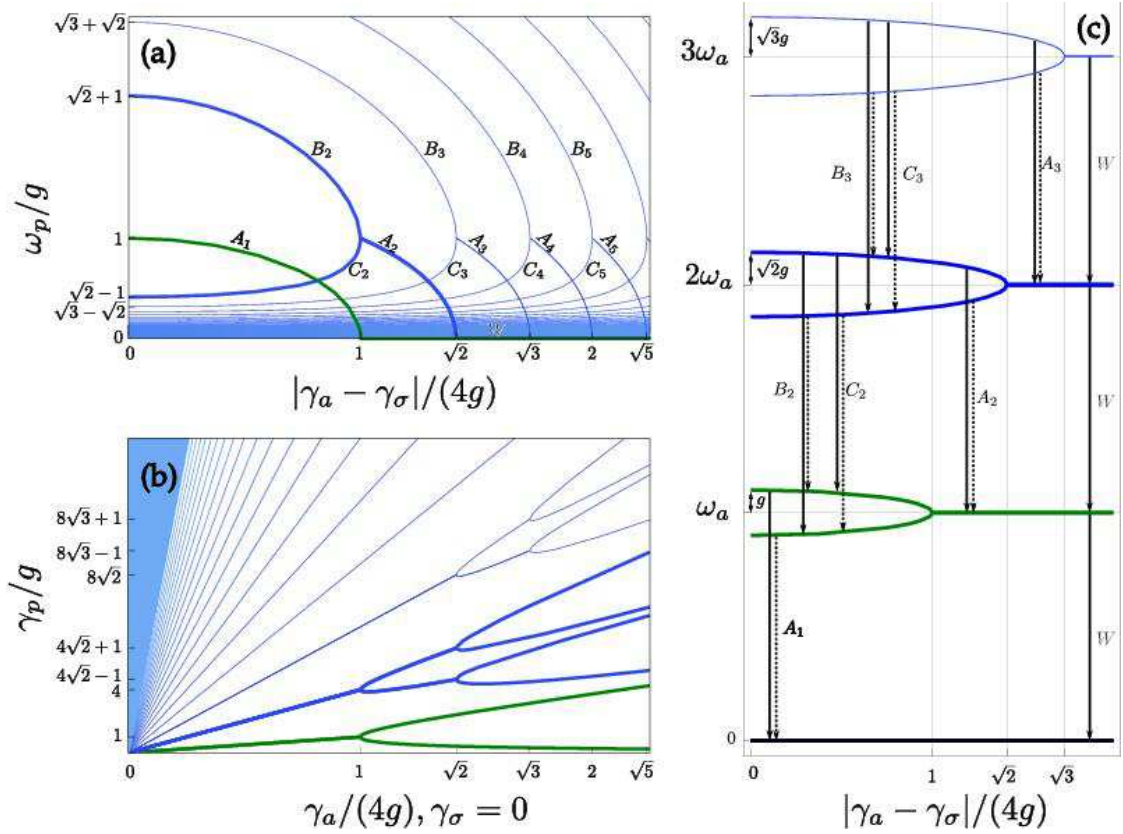}
  \caption{(Color online) Spectral structure of the Jaynes-Cummings
    model at resonance and without pumping. (a) Positions~$\omega_p$
    of the lines in the luminescence spectrum. Only energies higher
    than~$\omega_a$ are shown (not their symmetric
    below~$\omega_a$). We take~$\omega_a=0$ as the reference
    energy. In green (thick), the first manifold, and in increasing
    shades of blue, the successive higher manifolds which form a
    pattern of branch-coupling curves that define different orders of
    SC. (b) Half-Width at Half Maximum (HWHM)~$\gamma_p$ of the lines
    (with $\gamma_\sigma=0$). In both (a) and~(b), the blue filled
    region results from the accumulation of the countable-infinite
    vanishing lines. (c) Eigenenergies of the Jaynes-Cummings
    Hamiltonian with decay as an imaginary part of the bare energies
    (the \emph{Jaynes-Cummings ladder}). This provides a clear
    physical picture of panel~(a) where the peaks positions arise from
    the difference of energy between lines of two successive
    manifolds. Lines~$A_k$ of~(a) stem from the emission from
    manifold~$k$ in SC into manifold~$k-1$ in WC (or vacuum). Lines
    $B_k$ and~$C_k$ stem from the emission between the two manifolds
    in SC. Solid lines are those plotted in~(a), dotted lines produce
    the symmetric lines, not shown. The horizontal line~$W$ at~0 in
    (a) arises from decay between two manifolds in WC. Scheme~(c) also
    reproduces the broadening of the lines~(b) with the sum of the
    imaginary parts of the eigenenergies involved in the transition.}
  \label{fig:TueAug12144011CEST2008}
\end{figure}

In this Section, we discuss the series of coefficients~$\omega_p$
and~$\gamma_p$ that in the luminescence spectrum,
Eq.~(\ref{eq:FriOct26050345UTC2007}), determine the position and the
broadening (Half-Width at Half Maximum, HWHM) of the lines,
respectively, be it the cavity or direct exciton emission. The case of
vanishing pumping is fundamental, as it corresponds to the textbook
Jaynes-Cummings results with the spontaneous emission of an initial
state. It serves as the skeleton for the general case with arbitrary
pumping and supports the general physical picture. Finally, it admits
analytical results. We therefore begin with the case
where~$P_{a},P_\sigma\ll\gamma_{a},\gamma_\sigma$.  The eigenvalues of
the matrix of regression~$\mathbf{M}$, are grouped into
manifolds. There are two for the first manifold, given by:
\begin{equation}
  \label{eq:TueAug12190002CEST2008}
  D_{\substack{1\\2}}=\Gamma_++i\Big(\omega_a-\frac\Delta2\mp \sqrt{g^2-\Big(\Gamma_-+i\frac\Delta2\Big)^2} \Big)\,,
\end{equation}
and four for each manifold of higher order $k>1$, given by, for
$4k-5\le p\le 4k-2$:
\begin{equation}
  \label{eq:TueAug12185731CEST2008}
  D_p=\Gamma_k+i \Big( \omega_a+ \mathrm{sgn}\big({p-(8k-7)/2}\big) R_k+(-1)^pR^*_{k-1}\Big)\,,
\end{equation}
($\mathrm{sgn}(x)$ is defined as~$0$ for~$x=0$ and $x/|x|$ otherwise),
in terms of the $k$th-\emph{manifold (half) Rabi splitting}:
\begin{equation}
  \label{eq:ThuAug14163258CEST2008}
    R_k=\sqrt{(\sqrt{k}g)^2-\Big(\Gamma_-+i\frac\Delta2\Big)^2}\,,
\end{equation}
and of the $k$th-\emph{manifold (half) broadening}:
\begin{equation}
  \label{eq:TueAug12201052CEST2008}
  \Gamma_k=(2k-3)\Gamma_-+(2k-1)\Gamma_+=(k-1)\gamma_a+\frac{\gamma_\sigma}2\,.
\end{equation}
For each manifold, we have defined the~$D_p$ in order by increasing
value of the line position~$\omega_p$.

According to Eq.~(\ref{eq:SatAug2111206CEST2008}), these provide the
position~$\omega_p$ of the line and its half-broadening~$\gamma_p$
through their imaginary and real parts.  $\Gamma_k$ is always real, so
contributes in all cases to~$\gamma_p$ only. $R_k$ is (at resonance)
either pure real, or pure imaginary, and similarly to the boson case,
this is what defines SC.  This corresponds to an oscillatory or damped
field dynamics of the two-time correlators within manifold~$k$, which
lead us to the formal definition: WC and SC of order~$n$ are defined
as the regime where the complex Rabi frequency at resonance,
Eq.~(\ref{eq:ThuAug14163258CEST2008}), is pure imaginary (WC) or real
(SC).  The criterion for~$n$th order SC is therefore:
\begin{equation}
  \label{eq:ThuAug14131454CEST2008}
  g>|\Gamma_-|/\sqrt{n}\,.
\end{equation}

SC is achieve more easily for given system parameters ($g$
and~$\gamma_{a,\sigma}$), with an increasing photon-field intensity
that enhances the effective coupling strength. The lower the SC order,
the stronger the coupling.  This corresponds to the $n$th manifold
(and all above) being in SC (aided by the cavity photons), while the
$n-1$ manifolds below are in WC. First order is therefore the one
where all manifolds are in SC.
Equation~(\ref{eq:ThuAug14131454CEST2008}) includes the SC in the
standard boson case~\cite{elena_laussy08a}, $g>|\Gamma_-|$, as the
first order SC of the fermion case, that is shown in green (thick) in
Fig. \ref{fig:TueAug12144011CEST2008}. The same position of the
peaks~$\omega_{1,2}$ and the same (half) broadenings~$\gamma_{1,2}$ is
also recovered (in the absence of pumping).  Note that similarly to
the boson case, the SC is defined by a comparison between the coupling
strenght~$g$ with the \emph{difference} of the effective
broadening~$\Gamma_a$ and~$\Gamma_b$. The sum of these play no role in
this regard.

The $\omega_p$ and~$\gamma_p$ are plotted in
Fig.~\ref{fig:TueAug12144011CEST2008}(a) and
\ref{fig:TueAug12144011CEST2008}(b), respectively, as function
of~$\Gamma_-$. Note that~$\omega_p$ only depends on~$g$ and
$\Gamma_-$, whereas~$\gamma_p$ also depends on~$\Gamma_+$ (that is why
we plot it for $\gamma_\sigma=0$).

The $D_p$, Eq.~(\ref{eq:TueAug12185731CEST2008}), have a natural
interpretation in terms of transitions between the manifolds of the
so-called \emph{Jaynes-Cummings ladder}. The eigenenergies of the
Jaynes-Cummings Hamiltonian with decay granted as the imaginary part
of the bare energies~($\omega_{a,\sigma}-i\gamma_{a,\sigma}/2$), are
given by $E_{\pm}^k$ with
\begin{equation}
  E_{\pm}^k= k\omega_a-\frac{\Delta}2 \pm R_k -i\frac{(2k-1)\gamma_a+\gamma_b}{4}\,,
\end{equation}
for the $k$th manifold. The four possible transitions between
consecutive manifolds $k$ and $k-1$ give rise, when~$k>1$, to the four
peaks we found:
\begin{subequations}
  \label{eq:MonNov24151524GMT2008}
  \begin{align}
    \label{eq:TueNov25092036GMT2008}
    &D_{4k-5}=i[E^k_--(E^{k-1}_+)^*]\, ,\quad  &D_{4k-4}=i[E^k_--(E^{k-1}_-)^*]\,,\\
    \label{eq:TueNov25092101GMT2008}
    &D_{4k-3}=i[E^k_+-(E^{k-1}_+)^*]\, ,\quad
    &D_{4k-2}=i[E^k_+-(E^{k-1}_-)^*]\,.
  \end{align}
\end{subequations}
In the case $k=1$, only the two peaks common with the linear regime
arise, $D_{1,2}=iE^1_{\mp}$, given respectively by
Eqs.~(\ref{eq:TueNov25092036GMT2008})
and~(\ref{eq:TueNov25092101GMT2008}) with $E^0=0$. The fact that the
$D_p$ correspond to $i[E^k-(E^{k-1})^*]$ shows that, although the
positions of the lines are given by a difference, their broadenings
are given by a sum (because of complex conjugation). Physically, the
uncertainties in the initial and final states indeed add up in the
uncertainty of the transition energy.

The ladder is shown (at resonance) in
Fig.~\ref{fig:TueAug12144011CEST2008}(c). Let us discuss it in
connection with our definition of SC in this system, to arbitrary~$n$.
When $\Gamma_-=0$, each step of the ladder is constituted by the two
eigenstates of the Fermion dressed by the~$n$ cavity photons,
resulting in a splitting of~$2\sqrt{n}g$. This $n$-dependent splitting
produces quadruplets of delta peaks with splitting
of~$\pm(\sqrt{n}\pm\sqrt{n-1})g$ around~$\omega_a$, as opposed to the
boson case where independently of the manifold, the peaks are all
placed at~$\pm g$ around~$\omega_a$. In a more general situation with
$\Gamma_-\neq0$, there are three possibilities for a manifold~$k>1$:
\begin{enumerate}
\item \emph{Both manifold~$k$ and~$k-1$ are in SC.}  The two Rabi
  coefficients $R_k$ and~$R_{k-1}$ are real. This is the case when
  \begin{equation}
    \label{eq:ThuAug14145648CEST20081}
    |\Gamma_-|\le g\sqrt{k-1}\,.
  \end{equation}
  The luminescence spectra corresponds to four splitted
  lines~$\omega_p\rightarrow\omega_a\pm(R_k\pm R_{k-1})$, coming from
  the four possible transitions
  [Eqs.~(\ref{eq:MonNov24151524GMT2008}), shown as $B_k$ and~$C_k$ in
  Fig.~\ref{fig:TueAug12144011CEST2008}(c)] between manifolds~$k$
  and~$k-1$. The emission from all the higher manifolds also produces
  four lines. They are grouped pairwise around~$\omega_a$
  [Fig.~\ref{fig:TueAug12144011CEST2008}(a)] and all have the same
  broadening, contributed by~$\Gamma_k$ only [the single straight line
  in Fig.~\ref{fig:TueAug12144011CEST2008}(b)].
\item \emph{ Manifold~$k$ is in SC while manifold~$k-1$ is in WC.}  In
  this case, $R_k$ is pure imaginary (contributing to line positions)
  and~$R_{k-1}$ is real (contributing to broadenings).  This is the
  case when
  \begin{equation}
    \label{eq:SunNov23185732GMT2008}
    g\sqrt{k-1}<|\Gamma_-|<g\sqrt{k}\,.
  \end{equation}
  This corresponds to two lines~$\omega_p\rightarrow\omega_a\pm R_k$
  in the luminescence spectrum, coming from the two possible
  transitions [shown as $A_k$ in
  Fig.~\ref{fig:TueAug12144011CEST2008}(c)] between the SC
  manifold~$k$ and the WC manifold~$k-1$. Each of them is doubly
  degenerated. The two contributions at a given~$\omega_p$ have two
  distinct broadenings $\gamma_p\rightarrow\Gamma_k\pm |R_{k-1}|$
  around~$\Gamma_k$. [cf.~Fig.~\ref{fig:TueAug12144011CEST2008}(b)]. The
  final lineshapes of the two lines~$A_2$ is the same. In this region,
  all the emission from the higher manifolds produce four lines and
  all from the lower produce only one (at~$\omega_a$), being in WC.
\item \emph{Both manifold~$k$ and~$k-1$ are in WC}. The two Rabi
  coefficients $R_k$ and~$R_{k-1}$ are pure imaginary. This is the
  case when
  \begin{equation}
    \label{eq:ThuAug14145648CEST2008}
    g\sqrt{k}\le|\Gamma_-|\,.
  \end{equation}
  This corresponds to only one line at~$\omega_p\rightarrow\omega_a$
  in the luminescence spectrum, coming from the transition from one
  manifold in WC to the other [shown as $W$ in
  Fig.~\ref{fig:TueAug12144011CEST2008}(c)].  The line is four-time
  degenerated, with four contributions with different
  broadenings~$\gamma_p\rightarrow\Gamma_{k}\pm(|R_k|\pm |R_{k-1}|)$,
  as seen in Fig.~\ref{fig:TueAug12144011CEST2008}(b).
\end{enumerate}

Figure~\ref{fig:TueAug12144011CEST2008} is the skeleton for the
luminescence spectra---whether that of the cavity or of the direct
exciton emission. It specifies at what energies can be the possible
lines that constitutes the final lineshape, and what are their
broadening. To compose the final result, we only require to know the
weight of each of these lines.

In the SE case, the weights $L_p$ and $K_p$ include the integral of
the single-time mean values $\mathbf{v}_{a}(t,t)$ over $0\leq
t<\infty$. Therefore, only those manifolds with a smaller number of
excitations than the initial state can appear in the spectrum. Each of
them, will be weighted by the specific dynamics of the system. The
``spectral structure''---i.e., the~$\omega_p$ and~$\gamma_p$---depends
only the system parameters ($g$ and~$\gamma_{a,\sigma}$). Therefore,
in the SE case, the resulting emission spectrum is an exact mapping of
the spectral structure of the Hamiltonian,
Fig.~\ref{fig:TueAug12144011CEST2008}.

In the SS case, the weighting of the lines also depends on which
quantum state is realized, this time under the balance of pumping and
decay. But the excitation scheme also changes the spectral structure
of Fig.~\ref{fig:TueAug12144011CEST2008}. When the pumping parameters
are small, the changes will mainly be perturbations of the present
picture and most concepts will still hold, such as the definition of
SC, Eq.~(\ref{eq:ThuAug14131454CEST2008}) for nonzero~$P_{a,\sigma}$
in~(\ref{eq:TueAug12190422CEST2008}). However, when the pump
parameters are comparable to the decay parameters, the manifold
picture in terms of Hamiltonian eigenenergies breaks. The underlying
spectral structure must be computed numerically for each specific
probing of the system with~$P_a$ and~$P_\sigma$. It can still be
possible to identify the origin of the lines with the manifold
transitions by plotting their position~$\omega_p$ as a function of the
pumps, starting from the analytic limit. SC of each manifold can be
associated to the existence of peaks positioned at $\omega_p\neq
\omega_a,\omega_\sigma$. We address this problem in next Sections.

\section{Population and Statistics}
\label{TueAug12174728CEST2008}
 
To know which features of the spectral structure dominate and which
are negligible, one needs to know what is the quantum state of the
system. In the boson case, it was enough to know the average photon
($n_a$) and exciton ($n_b$) numbers, and the off-diagonal
element~$n_{ab}=\langle\ud{a}b\rangle$. In the most general case of
the fermion system, a countably infinite number of parameters are
required for the exact lineshape. The new order of complexity brought
by the fermion system is illustrated for even the simplest
observable. Instead of a closed relationship that provides, e.g., the
populations in terms of the system parameters and pumping rates, only
relations between observables can be obtained in the general case. For
instance, for the populations:
\begin{equation}
  \label{eq:ThuAug14203427CEST2008}
  \Gamma_a n_a+\Gamma_\sigma n_\sigma=P_a+P_\sigma\,.
\end{equation}
This expression is formally the same as for the coupling of two
bosonic modes. The differences are in the effective dissipation
parameter $\Gamma_\sigma=\gamma_\sigma+P_\sigma$ (instead of the
bosonic one, $\gamma_b-P_b$) and the constrain of the exciton
population, $0\leq n_\sigma\leq 1$. One solution of
Eq.~(\ref{eq:ThuAug14203427CEST2008})
is~$n_a^\mathrm{th}=P_a/\Gamma_a$ and
$n_\sigma^\mathrm{th}=P_\sigma/\Gamma_\sigma$, which corresponds to
the case $g=0$, where each mode reaches its thermal steady state
independently (Bose/Fermi distributions, depending on the mode
statistics). With coupling~$g\neq0$, we can only derive some
analytical limits and bounds. For example, when $\gamma_a=P_a$, one
sees from Eq.~(\ref{eq:ThuAug14203427CEST2008}) that
$n_\sigma=(P_\sigma+P_a)/\Gamma_\sigma$, with the condition for the
cavity pump~$P_a\leq\gamma_\sigma$ (since $n_\sigma\leq 1$). If only
the dot is pumped, $n_\sigma=n_\sigma^\mathrm{th}$, and if both $P_a$,
$\gamma_a=0$ then, also
$n_a=n_a^\mathrm{th}=P_\sigma/(\gamma_\sigma-P_\sigma)$ with the same
temperature. As, in this case, $P_\sigma$ must be strictly smaller
than~$\gamma_\sigma$, the exciton population $n_\sigma\leq 1/2$
prevents an inversion of population, as is well known for a two-level
system.

When $\gamma_a>P_a$, we get the following bounds for the cavity
populations in terms of the system and pumping parameters:
\begin{equation}
  \label{eq:ThuAug14203713CEST2008}
  \frac{P_a-\gamma_\sigma}{\gamma_a-P_a}\le n_a\le\frac{P_a+P_\sigma}{\gamma_a-P_a}\,.
\end{equation}
When $P_\sigma=\gamma_\sigma=0$, the cavity is in thermal equilibrium
with its bath, $n_a=n_a^\mathrm{th}$, and with the dot
$n_\sigma=P_a/(\gamma_a+P_a)$. In this case, the pump is limited by
$P_a<\gamma_a$, and again $n_\sigma\le1/2$. Again, the inversion of
population cannot take place putting the system in contact with only
one thermal bath.  In all these situations where an analytic
expression for the population is obtained, the detuning between cavity
and dot does not affect the final steady state, although it
determines, together with the coupling strength, the time that it
takes to reach it. An interesting limiting case where inversion can
happen, is that where $\gamma_\sigma$ and $P_a$ are negligible, then
$n_a=P_\sigma(1-n_\sigma)/\gamma_a$. When the pump is low and
$n_\sigma< 1$, $n_a$ grows with pumping, but when the dot starts to
saturate and $n_\sigma\rightarrow 1$ the cavity population starts to
quench towards~$n_a\rightarrow0$~\cite{benson99a}. Here, all values
of~$P_\sigma$ bring the system into a steady state as $n_a$ cannot
diverge. However, if we allow some cavity pumping, given that~$a$ does
not saturate, $P_a$ is bounded. A rough guess of this boundary is, in
the most general case:
\begin{equation}
  \label{eq:ThuAug14204909CEST2008}
  P_a<\max(\gamma_a,\gamma_\sigma)\,.
\end{equation}
If Eq.~(\ref{eq:ThuAug14204909CEST2008}) is not fulfilled, the system
diverges, as more particles are injected at all times by the
incoherent cavity pumping than are lost by decay. Numerical evidence
suggests that the actual maximum value of~$P_a$ depends on
$P_\sigma$. To some given order~$n_t$, divergence typically arises
much before condition~(\ref{eq:ThuAug14204909CEST2008}) is reached,
although it is difficult to know if a lower physical limit has been
reached or if the order of truncation was not high enough.

The second order correlator~$g^{(2)}$ can be expressed as a function
of~$n_a$ only:
\begin{multline}
  \label{eq:MonJul21005054CEST2008}
  g^{(2)}=\Big[g^2\big((n_a+1)(P_a^2+P_\sigma^2)-n_a(\gamma_a+\gamma_\sigma)^2+P_a(\gamma_a+\Gamma_\sigma+P_\sigma+6n_a\Gamma_\sigma)\\
  {}+P_\sigma(\gamma_a+\gamma_\sigma-2n_a\gamma_a)\big)+(P_a-n_a\Gamma_a)\Gamma_\sigma(4\Gamma_+^2+\Delta^2)\Big]\Big/2g^2n_a^2\Gamma_a\Gamma_\sigma\,.
\end{multline}
Obtaining the expression for the~$n$th order correlator and setting it
to zero would provide an approximate (of order~$n$) closed relation
for~$n_a$. We shall not pursue this line of research that becomes very
heavy.

\begin{figure}[hbtp]
  \centering
  \includegraphics[width=.75\linewidth]{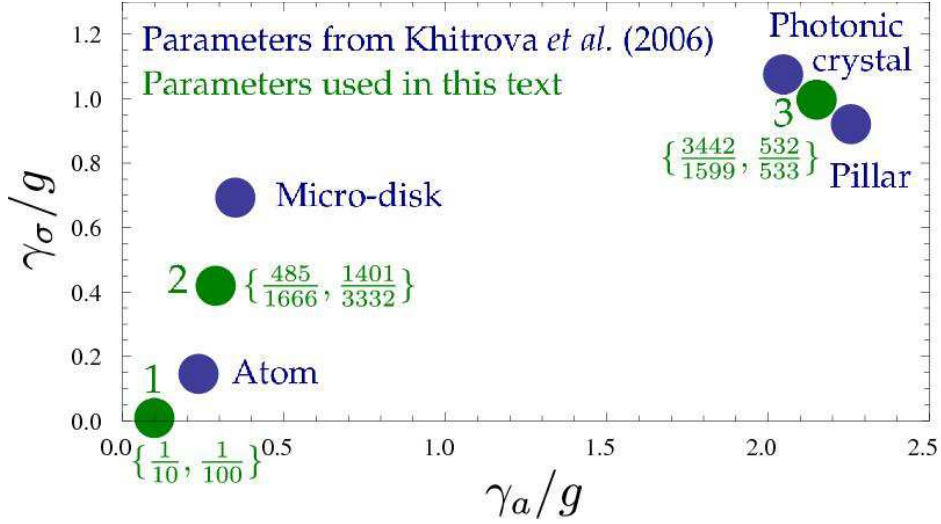}
  \caption{(Colour online) Blue points give the decay rates for the
    cavity and quantum dot estimated by Khitrova \emph{et al.} in
    Ref.~\cite{khitrova06a} for four references systems having
    achieved SC at this time: photonic crystals and pillar
    microcavities nearby point~3, microdisks and atomic systems nearby
    point~2. In Green, the three sets of parameters used in this
    text. Points~2 and 3 average over their two nearest neighbors and
    represent these systems.  Point~1 represents a very good system in
    very strong coupling, that might be realizable in the near
    future. Parameters are fractions because numerical computations
    have been done to arbitrary precisions (with the values given).}
  \label{TueSep16163717BST2008}
\end{figure}

\begin{figure}[hbpt]
  \centering
    \includegraphics[width=\linewidth]{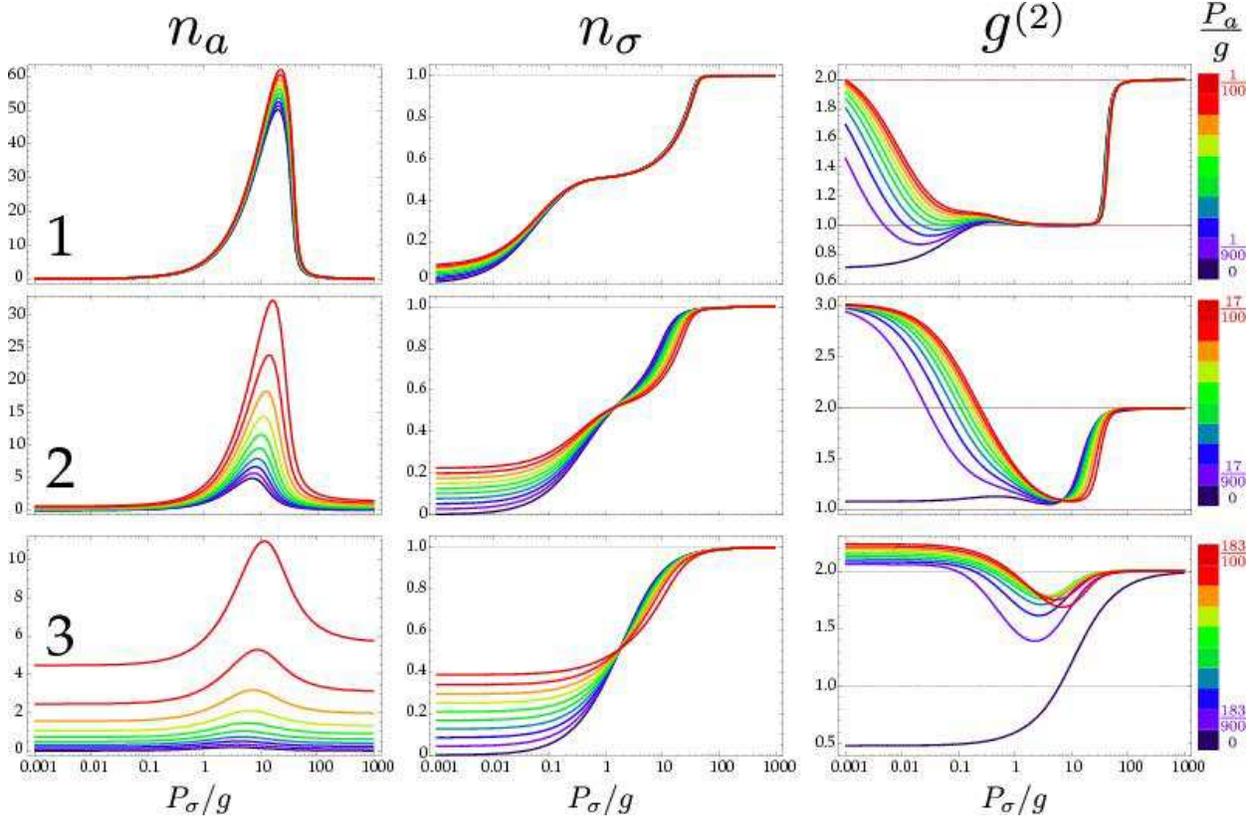}
    \caption{Populations and statistics of the points marked 1, 2
      and~3 in Fig.~\ref{TueSep16163717BST2008}. Each row shows the
      triplet $n_a$ (1st column), $n_\sigma$ (2nd) and~$g^{(2)}(0)$
      (3rd) for a given point ($n$th row corresponds to
      point~$n$). All plots share the same $x$-axis in log-scale
      of~$P_\sigma/g$ ranging from $10^{-3}$ to $10^3$. All $y$-axis
      are rescaled to its specific graph, at the exception
      of~$n_\sigma$ which is always between~0 and~1. The color code
      corresponds to different values of~$P_a$. Each color code
      applies to its row and is given in the last column. The
      qualitative behavior is roughly the same for all points: there
      is a peak in~$n_a$ that is subsequently quenched as the dot gets
      saturated. In $g^{(2)}$, there is on the other hand, a local
      minimum of fluctuations that can be brought to the Poissonian
      limit of~$1$ (allowing for a lasing region) and maintained over
      a large plateau in good cavities.}
  \label{fig:SunOct5183310BST2008}
\end{figure}

As an overall representation of the typical systems that arise in real
and desired experiments, we consider three configurations, shown in
Fig.~(\ref{TueSep16163717BST2008}), scattered in order to give a rough
representative picture of the overall possibilities, around parameters
estimated in Ref.~\cite{khitrova06a}. Point~1 corresponds to the best
system of our selection, in the sense that its decay rates are very
small ($\gamma_a=g/10$, $\gamma_\sigma=g/100$), and the quantum
(Hamiltonian) dynamics dominates largely the system. It is a system
still outside of the experimental reach. Point~3 on the other hand
corresponds to a cavity with important dissipations, that, following
our analysis below, precludes the observation of any neat structure
attributable to the underlying Fermi statistics. According to
numerical fitting of the experiment, real structures might even be
suffering higher dissipation rates~\cite{laussy08a}.  Point~2
represents other lead systems of the SC physics, that we will show can
presents strong departure from the linear regime, in particular
conditions that we will emphasize.  The best semiconductor system from
Fig.~\ref{TueSep16163717BST2008} is realized with microdisks, thanks
to the exceedingly good cavity factors. We shall not enter into
specific discussion of the advantages and inconvenient of the
respective realizations and the accuracy of these estimations. From
now on, we shall refer to this set of parameters as that of
``reference points'', keeping in mind that points~1 and~2 in
particular represent systems that we will refer to as a ``good
system'' and a ``more realistic system'', respectively.

In Fig.~\ref{fig:SunOct5183310BST2008}, the three observable of main
interest for a physical understanding of the system that we have just
discussed---$n_a$, $n_\sigma$ and~$g^{(2)}$---are obtained numerically
for the three reference points.  Electronic pumping is varied from,
for all practical purposes, vanishing ($10^{-3}g$) to infinite
($10^3g$) values. Various cavity pumpings are investigated and
represented by the color code from no-cavity pumping (dark blue) to
high, near diverging, cavity pumping (red), through the color
spectrum. We checked numerically that these results
satisfy~Eq.~(\ref{eq:ThuAug14203427CEST2008}).  The overall behavior
is mainly known, for instance the characteristic increase till a
maximum and subsequent decrease of $n_a$ with~$P_\sigma$ has been
predicted in a system of QD coupled to a
microsphere~\cite{benson99a}. This phenomenon of so-called
\emph{self-quenching} is due to the excitation impairing the coherent
coupling of the dot with the cavity: bringing in an exciton too early
disrupts the interaction between the exciton-photon pair formed from
the previous exciton. Therefore the pumping rate should not overcome
significantly the coherent dynamics. Too high electronic pumping
forces the QD to remain in its excited state and thereby prevents it
from populating the cavity. In this case the cavity population returns
to zero while the exciton population (or probability for the QD to be
excited) is forced to one. The cavity pumping brings an interesting
extension to this mechanism. First there is no quenching for the
pumping of bosons that, on the contrary, have a natural tendency to
accumulate and lead to a divergence. Therefore the limiting values
for~$n_a$ when~$P_\sigma\rightarrow0$ or~$P_\sigma\rightarrow\infty$
are not zero, as in the previously reported self-quenching
scenario~\cite{benson99a}.  They also happen to be different:
\begin{subequations}
  \label{eq:ThuNov6235857GMT2008}
  \begin{align}
    n_a^<&\equiv n_a(P_\sigma=0)=\frac{P_a-\gamma_\sigma n_\sigma}{\gamma_a-P_a}\,,\\
    n_a^>&\equiv\lim_{P_\sigma\rightarrow\infty}n_a=\frac{P_a}{\gamma_a-P_a}\,,\label{eq:FriNov7111433GMT2008}
  \end{align}
\end{subequations}
and therefore satisfy~$n_a^<<n_a^>$.
Eq.~(\ref{eq:FriNov7111433GMT2008}) follows from the decoupled thermal
values for the populations, $n_\sigma\rightarrow
P_\sigma/(P_\sigma+\gamma_\sigma)$, and corresponds to a passive
cavity where the quenched dot does not contribute at all. In this
case, the emission spectrum of the system is expected to converge to
\begin{equation}
  \label{eq:WedOct29092210GMT2008}
  S_a(\omega)=\frac1\pi\frac{\Gamma_a/2}{(\omega-\omega_a)^2+(\Gamma_a/2)^2}\,,
\end{equation}
for the cavity, and~$S_\sigma(\omega)=0$ for the dot. The other limit
when~$P_\sigma=0$, shows the deleterious effect of the dot on cavity
population. The dot fully enters the dynamics contrary to the quenched
case where it is subtracted from it.

Important application of SC for single-atom lasing are to be found in
good cavities 1 and~2, where the coupling $g\gg\gamma_a,\gamma_b$ is
strong
enough~\cite{mu92a,gincel93a,jones99a,karlovich01a,kozlovskii99a}. Lasing
can occur when the pumping is also large enough to overcome the total
losses, $P_\sigma\gg\gamma_a,\gamma_\sigma$. Setting $P_a$,
$\gamma_\sigma=0$, Eqs.~(\ref{eq:SunJul20204529CEST2008}) can be
approximately reduced to one for the total probability $\mathrm{p}[n]$
~\cite{scully_book02a,benson99a}:
\begin{equation}
  \label{eq:WedDec10105804GMT2008}
  \partial_t\mathrm{p}[n]=\gamma_a(n+1)\mathrm{p}[n+1]-\Big(\gamma_an+\frac{l_G(n+1)}{1+l_S/l_G(n+1)}\Big)\mathrm{p}[n]+\frac{l_Gn}{1+l_S/l_Gn}\mathrm{p}[n-1]
\end{equation}
The parameters that characterize the laser are the gain
$l_G=4g^2/P_\sigma$ and the self saturation
$l_S=8g^2l_G/P_\sigma^2$. Far above threshold ($n_al_S/l_G\gg1$), the
statistics are Poissonian, $g^{(2)}=1$, with a large intensity in the
emission, $n_a=P_\sigma/(2\gamma_a)$, and half filling of the dot,
$n_\sigma=0.5$. However, this analytic limit from the standard laser
theory is not able to reproduce the self-quenching effect induced by
the incoherent pump, nor the subpoissonian region ($g^{(2)}<1$) where
quantum effects are prone to appear. The validity of the laser theory
is restricted to the narrow region, $\gamma_a\ll P_\sigma\ll
\gamma_\sigma^\mathrm{P}$, where
$\gamma_\sigma^\mathrm{P}=4g^2/\gamma_a$ is the boundary for the
self-quenching. In the weak coupling regime,
$\gamma_\sigma^\mathrm{P}$ is the well known Purcell enhanced
spontaneous decay rate of an exciton through the cavity mode. In the
strongly coupled system, it can be similarly understood as the rate at
which an exciton transforms into a photon. If the excitons are
injected at a higher rate, there is no time for such a coherent
exchange to take place and populate the cavity with
photons. Fig.~\ref{fig:SunOct5183310BST2008} shows that lasing can be
achieved with system~1 in the corresponding region of pump.  Here, we
will solve the system exactly, covering this regime and all the other
possible ones with the full quantum equations of motion.

The effect of cavity pumping depends strongly on the experimental
situation. In the case of an exceedingly good system, $P_a$ has little
effect as soon as the exciton pumping is important,
$P_\sigma>\gamma_a$.  Cavity pumping becomes important again in a
system like~2, where it enhances significantly the output power, with
the price of superpoissonian statistics ($g^{(2)}>1$). With a poorer
system like point~3, some lasing effect can be found with the aid of
the cavity pump: there is a nonlinear increase of $n_a$ and $g^{(2)}$
approaches 1 for $g<P_\sigma<10g$. However, the weaker the coupling,
the weaker this effect until it disappears completely for decay rates
outside the range plotted in Fig.~\ref{TueSep16163717BST2008}. In all
cases, the self-quenching leads finally to a thermal mixture of
photons ($g^{(2)}=2$) and WC at large pumping.

\section{Weights and Renormalization}
\label{SunNov16135522GMT2008}

\begin{figure}[hbpt]
  \centering
  \includegraphics[width=.5\linewidth]{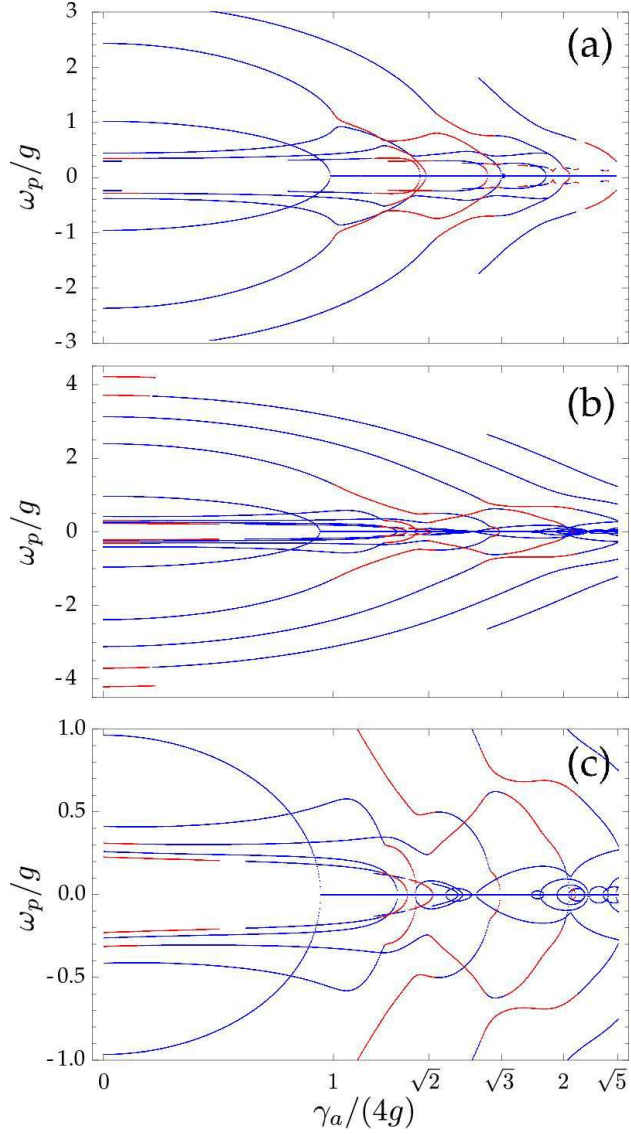}
  \caption{(Color online) Spectral structure in the cavity emission of
    the Jaynes-Cummings model as a function of $\gamma_a/g$, with some
    electronic pumping ($\Delta=0$, $\gamma_\sigma=0$, $P_a=0$,
    $\omega_a=0$). Panel (a) is for~$P_\sigma=g/50$ and (b)-(c)
    for~$P_\sigma=g/10$. Panel (c) is a zoom on the central peaks of
    the entire picture (b). In blue (resp., red) are the peaks
    with~$L_p^a>0$ (resp.,~$L_p^a<0$).}
  \label{fig:SunNov9170957GMT2008}
\end{figure}

\begin{figure}[htbp]
  \centering
  \includegraphics[width=\linewidth]{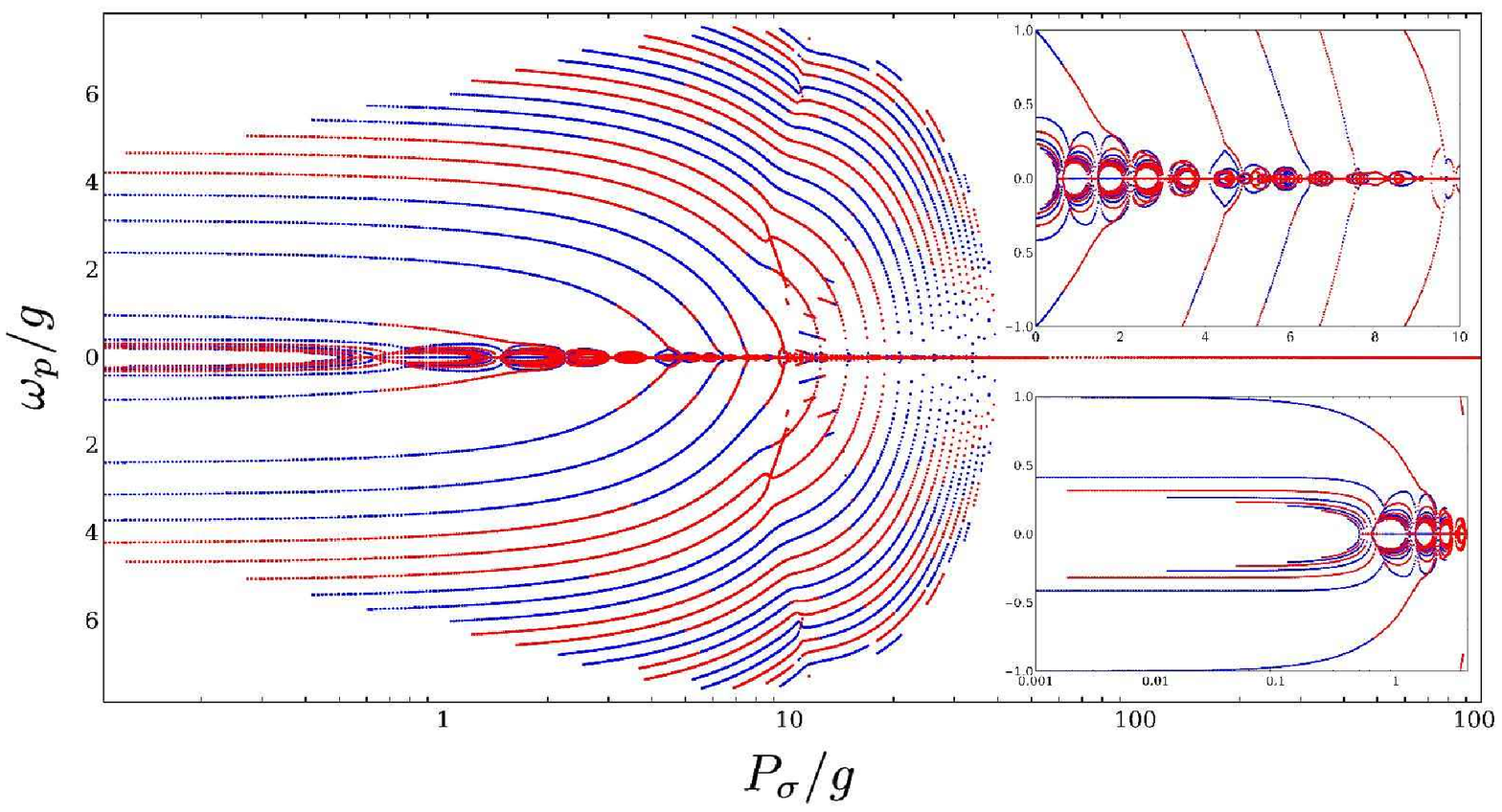}
  \caption{(Color online) Spectral structure in the cavity emission at
    resonance as a function of $P_\sigma/g$ for Point 2 ($\Delta=0$,
    $P_a=0$, $\omega_a=0$). In insets, we zoom over the central peaks
    and the region $0\le P_\sigma\le g$ (upper) and $0\le P_\sigma \le
    10g$ (lower), showing the complex structures that arise. In blue
    (resp., red) are the peaks with~$L_p^a>0$ (resp.,~$L_p^a<0$). At
    sufficiently high pumping, all eigenvalues have collapse to zero,
    defining an extreme case of \emph{weak coupling}.}
  \label{fig:TueNov11182731GMT2008}
\end{figure}

To give a complete picture of the spectral structure, that we have
obtained analytically in Section~\ref{TueAug12172114CEST2008}, we need
to consider how this limiting case of vanishing pumpings evolves with
finite pumping. Here again, we have to turn to numerical results. 

Two cases of finite pumpings are shown in
Fig.~\ref{fig:SunNov9170957GMT2008} for the finite pumping counterpart
of Fig.~\ref{fig:TueAug12144011CEST2008}(a), namely~$P_\sigma=g/50$,
(a), and~$P_\sigma=g/10$, (b) and~(c). We take $\omega_a=0$ as the
reference energy for the remaining of this text. Panel~(a) shows how
the limiting case ($P_\sigma \ll g$) is weighted and deviates rather
lightly from the analytical result. The computation has been made to
truncation order~$n_t=50$ and we checked that it had converged with
other truncation orders giving exactly the same result. In the figure,
only~$\omega_p$ whose weighting in the cavity emission~$L^a_p$
(Lorentzian part) is nonzero are shown, although most of them are very
small. If we plot only those with~$|L_p^a|\ge 0.01$, only the usual
\emph{vacuum Rabi doublet} (in green in
Fig.~\ref{fig:TueAug12144011CEST2008}) would remain. In addition of
the weight, also the degeneracy (number of peaks) at a given resonance
should be taken into account to quantify the intensity of emission at
a particular energy. This information is not apparent in the figures,
where we only show in Blue or Red the cases of positive or negative,
respectively, weighting. In some cases, many peaks superimpose with
opposite signs, possibly cancelling each other. We plot negative
values last so that a blue line corresponds to a region of only
positive values, while a red line may come on top of a blue line. This
figure gives nevertheless an insightful image of the underlying energy
structure and how they contribute to the final spectrum as an addition
of many emitting (or interfering) events. In~(b) we show a case of
higher pumping, with the same principal information to be found in the
mapping of the eigenvalues. The characteristic branch-coupling of the
Jaynes-Cummings, still easily identifiable in
Fig.~\ref{fig:SunNov9170957GMT2008}(a), has vanished, and lines of
external peaks directly collapse toward the center. A zoom of the
central part, panel~(c), shows the considerable complexity of the
inner peaks, forming ``bubbles'' around the central line, due to
intensity-aided SC fighting against increasing dissipation that
ultimately overtakes.

The origin of the lines can be better understood if we plot them as a
function of pumping, as we commented in
Section~\ref{TueAug12172114CEST2008}. In
Fig.~\ref{fig:TueNov11182731GMT2008}, the same weighted peak
positions~$\omega_p$ are shown (with the same color code) for Point~2
as electronic pumping is varied from~$10^{-3}g$ to~$10^3g$
($P_a=0$). This last picture supports the idea that quantum effects
(such as subpoissonian statistics,
Fig.~\ref{fig:SunOct5183310BST2008}) are observed at small pumpings,
with optimal range being roughly~$P_\sigma<0.5g$, where only the
lowest manifolds are probed. This is the range of pumping where the
Jaynes-Cummings manifold structure is still close to that without
pump. Further pumping pushes the lines to collapse, starting by the
vacuum Rabi splitting which closes, evidencing the loss of the first
order SC at $P_\sigma\approx4g$. Here again we observe this phenomenon
of bubbling, with a sequence of lines opening and collapsing, that
makes it impossible to specify the exact pump at which the transition
takes place. From this point, SC is lost manifold by manifold
similarly as in the case where $\gamma_a$ was
increased. When~$P_\sigma\approx 40g$, all lines have collapsed onto
the center and will remain so at higher pumpings. The dot saturates
and the cavity empties with thermal photons in a WC regime.

In these conditions, either from
Fig.~\ref{fig:SunNov9170957GMT2008}(b)-(c) or
Fig.~\ref{fig:TueNov11182731GMT2008}, a general definition of
Strong-Coupling in presence of pumping is obviously very complex and
remains to be established. 

\section{Luminescence Spectra at Resonance}
\label{SunNov16135934GMT2008}

Now we have all the ingredients to present the final result: the
spectral shapes of the system in a broad range of configurations and
parameters.  We cannot give a comprehensive picture as any set of
parameters is by itself unique, but will instead illustrate the main
trends, using specifically for that purposes the three reference points
of Fig.~\ref{TueSep16163717BST2008}. They give a good account of the
general case and one can extrapolate from these particular cases how
another configuration will behave. To get exact results for a given
point, numerical computations must be undertaken.

\begin{figure}[hbpt]
  \centering
  \includegraphics[width=.5\linewidth]{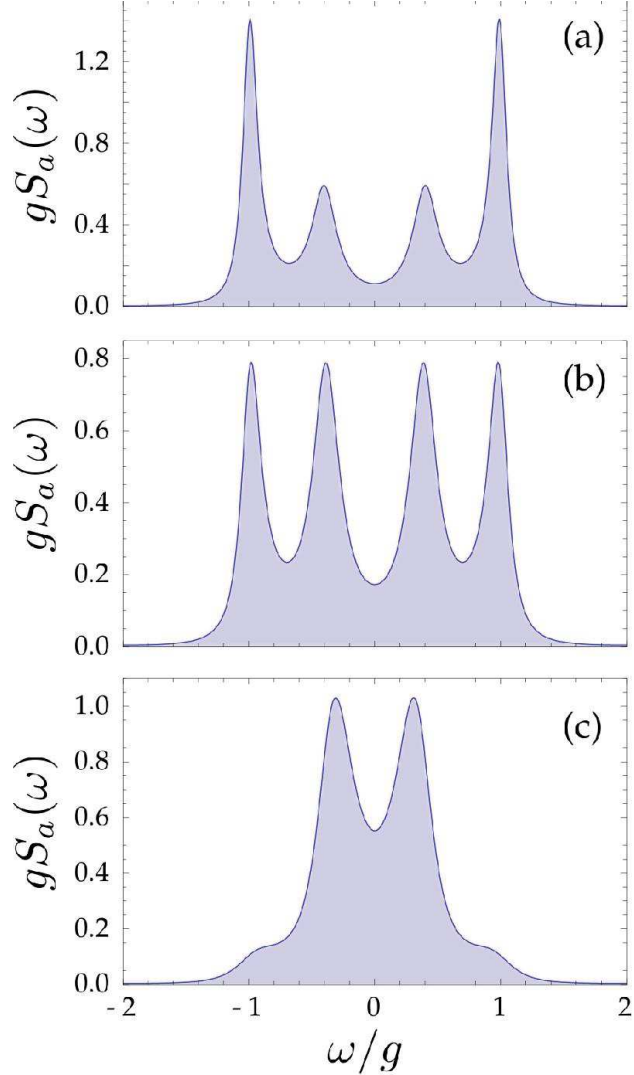}
  \caption{(Color online). Jaynes-Cummings forks as they appear in the
    luminescence spectrum of a QD in a microcavity with system
    parameters given by Point 1 of Fig.~\ref{TueSep16163717BST2008}
    and for pumping rates $(P_a,P_\sigma)/g$ given by (a),
    $(0,0.057)$; (b), $(0.002,0.087)$ and (c), $(0.001,0.27)$. The two
    outer peaks at~$\pm 1$ correspond to the vacuum Rabi doublet of
    the linear limit. Inner peaks correspond to transitions with
    states of more than one excitation. Although the underlying
    structure is the same, many variations of the actual lineshapes
    can be obtained.}
  \label{fig:ThuOct23180021BST2008}
\end{figure}

\begin{figure}[hbpt]
  \centering
  \includegraphics[width=.75\linewidth]{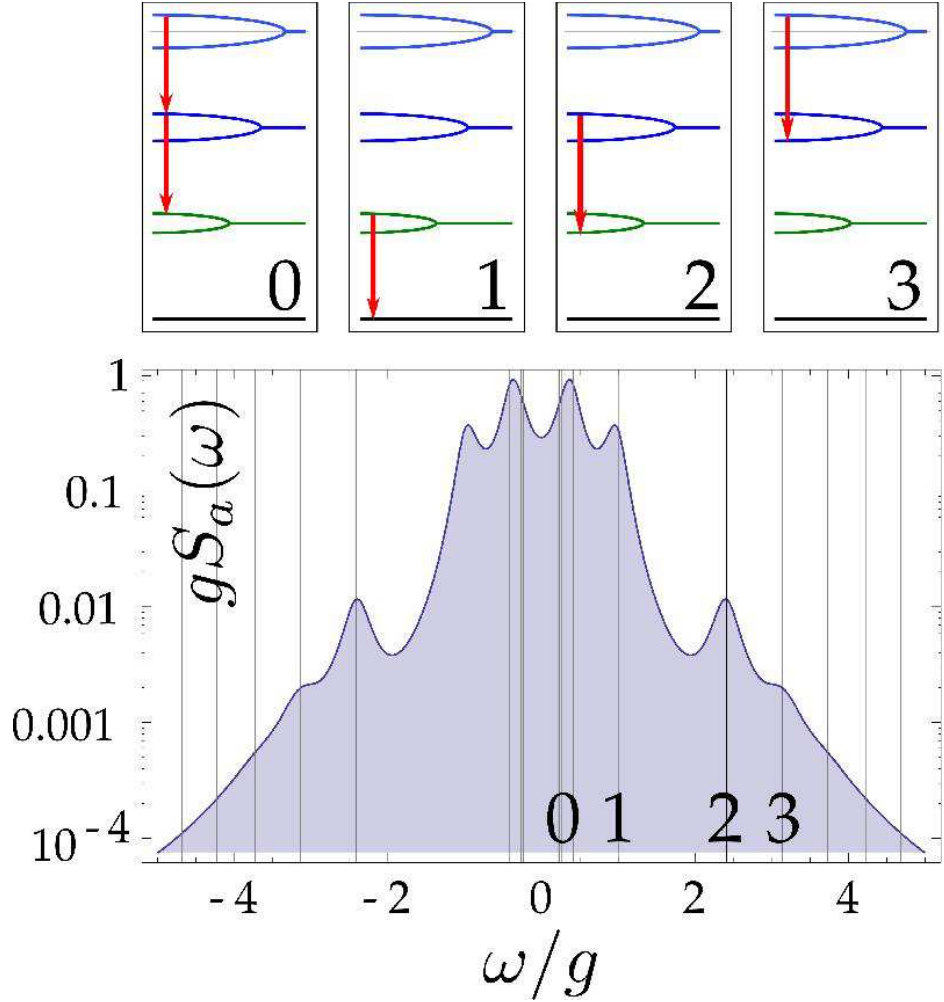}
  \caption{(Color online). Expanded view in logarithmic scale of a
    spectrum similar to those of Fig.~\ref{fig:ThuOct23180021BST2008},
    this time with $(P_a,P_\sigma)/g=(0.002,0.076)$. Transitions up to
    the third manifold (shown in insets) are resolvable. Others are
    lost in the broadening. The transition energies of the
    Jaynes-Cummings ladder are shown by vertical lines (up to the
    third manifold). The Rabi peaks that corresponds to transitions
    from the first manifold to vacuum (line~1) is in this case
    dominated by higher transitions that accumulate close to the
    center (line~0).}
  \label{fig:ThuOct23180022BST2008}
\end{figure}

\begin{turnpage}
  \begin{figure}[hbpt]
    \centering
    \includegraphics[width=\linewidth]{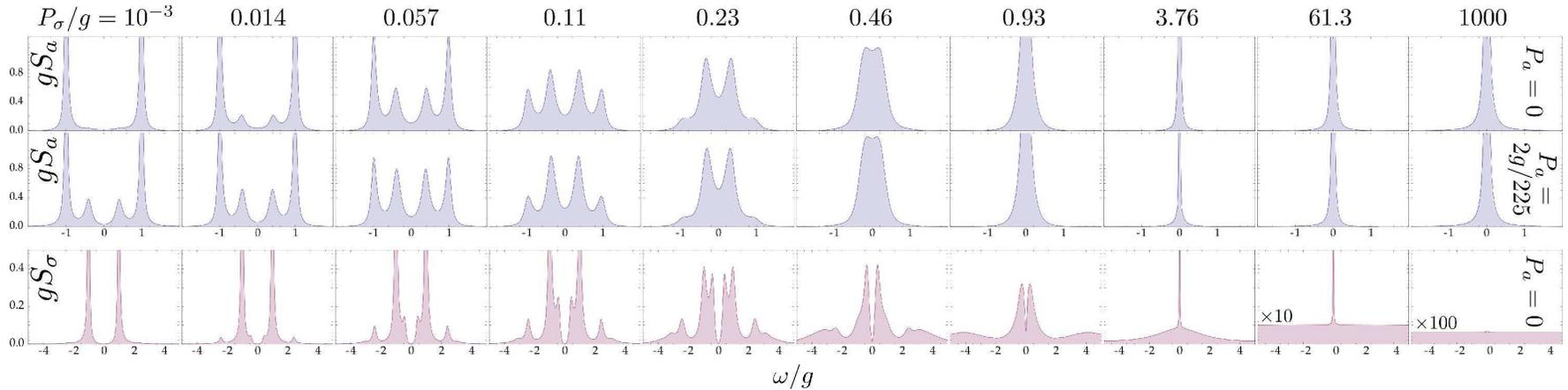}
    \caption{(Color online). Point 1 of
      Fig.~\ref{TueSep16163717BST2008}. Spectral emission over a wide
      range of electronic pumping~$P_\sigma/g$ from~$10^{-3}$
      to~$10^3$ showing the three main regimes: multiplet emission,
      lasing and quenching.  Cavity pumping only affects
      quantitatively the main features of the emission in this case of
      very strong coupling, so a small set is shown as representative
      enough: two upper rows (blue) correspond to cavity emission for
      no and large cavity pumping, respectively, and lower row
      (violet) to the direct exciton emission for no cavity
      pumping. The Jaynes-Cummings fork is clearly resolved at small
      $P_\sigma$ ($<0.2g$) and is enhanced by the cavity pumping. At
      higher electronic pumping ($P_\sigma\approx 0.5 g$), the
      multiplet structure collapses into a dominant doublet of inner
      peaks while the vacuum Rabi peaks melt into its shoulders. Then
      the system is brought into lasing ($g<P_\sigma<30g$) and is
      finally quenched ($P_\sigma>30g$).}
    \label{fig:SunOct19212128BST2008}
  \end{figure}
\end{turnpage}

From now on, we shall represent in Blue the spectra that correspond to
cavity emission and in Violet those that correspond to direct exciton
emission. The main conclusions, based on semi-analytical results, are
different for different points or family of points. Point 1, that
represents a very good system, is the one that is best suited to
explore quantum effects. Its spectral shape is unambiguously
evidencing transitions in the Jaynes-Cummings ladder, as shown in
Fig.~\ref{fig:ThuOct23180021BST2008} with a clear ``Jaynes-Cummings
fork'' (a quadruplet). The outer peaks at~$\pm 1$ are the conventional
vacuum Rabi doublet, whereas the two inner peaks correspond to higher
transitions in the ladder. Observation of a transition from outer to
inner peaks with pumping such as shown in
Fig.~\ref{fig:ThuOct23180021BST2008} would be a compelling evidence of
a quantum exciton in SC with the
cavity. Fig.~\ref{fig:ThuOct23180022BST2008} shows another multiplet
structure of this kind for Point~1. The intensity of emission is
presented in log-scale and for a broader range of frequencies, so that
small features can be revealed. Transitions from up to the third
manifold can be explicitly identified. The decay from the second
manifold, that manifests distinctly with peaks labelled 2 (although it
also contributes to peaks labelled 0), is already weak but still might
be identifiable in an experimental PL measurement. Higher transitions
have decreasing strenght. This can be checked with the probability to
have $n$ photons in the cavity, $\mathrm{p}(n)$, computed from
Eqs.~(\ref{eq:SunJul20204529CEST2008}). Whenever the mean number $n_a$
is low (as is the case here), this probability is maximum for the
vacuum ($\mathrm{p}(n)>\mathrm{p}(n+1)$ for all $n$), independently of
the nature of the photon distribution (sub, super or Poissonian). Only
when $n_a=1$, in the best of cases (for a Poissonian distribution),
does this trend start to invert and
$\mathrm{p}(1)=\mathrm{p}(0)$. This makes it impossible, even in the
very good system of Point 1, to probe clearly and independently
transitions between manifolds higher than~$3$, as their weak two outer
peaks (approximately at $\pm(\sqrt{n}+\sqrt{n-1})$) are completely
hidden by the broadening. A stronger manifestation of nonlinear
emission is to be found in the pool of pairs of inner peaks from all
high-manifold transitions (labelled 0 in
Fig.~\ref{fig:SunOct19212128BST2008}), at approximately $\pm
(\sqrt{n}-\sqrt{n-1})$. Not only the inner peaks coming from different
manifolds are close enough to sum up, but also they are more intense
than their outer counterparts. This can be easily understood by
looking at the probability, $\mathrm{I}_{c}$, of transition between
eigenstates $\ket{\pm,n}$ through the emission of a photon, $c=a$, or
an exciton, $c=\sigma$. This probability,
$\mathrm{I}_{c}^{(i\rightarrow f)}\propto|\bra{f}c\ket{i}|^2$,
estimates the relative intensity of the peaks depending on the
initial, $\ket{i}$, and final, $\ket{f}$, states of the transition and
on the channel of emission, $c=a,\sigma$~\cite{laussy06b}. A
discussion in terms of the eigenstates of the Hamiltonian is still
valid in the regime of Point~1 (very good system) at very low pump. At
resonance, neglecting pumps and decays, the eigenstates for manifold
$n$, are $\ket{n,\pm}=(\ket{n,0}\pm\ket{n-1,1})/\sqrt{2}$. The outer
peaks arise from transitions between eigenstates of different kind,
$\ket{n,\pm}\rightarrow\ket{n-1,\mp}$, while the inner peaks arise
from transitions between eigenstates of the same kind,
$\ket{n,\pm}\rightarrow\ket{n-1,\pm}$. Their probability amplitudes in
the cavity emission,
\begin{subequations}
  \label{eq:WedNov26131557GMT2008}
  \begin{align}
    \label{eq:WedNov26131626GMT2008}
    &\mathrm{I}_{a}^{(\pm\rightarrow\mp)}\propto|\bra{n-1,\mp}a\ket{n,\pm}|^2=|\sqrt{n}-\sqrt{n-1}|^2/4\,,\\
    &\mathrm{I}_{a}^{(\pm\rightarrow\pm)}\propto|\bra{n-1,\pm}a\ket{n,\pm}|^2=|\sqrt{n}+\sqrt{n-1}|^2/4\,,
  \end{align}
\end{subequations}
evidence the predominance of the inner peaks versus the outer ones,
given that one expect the same weighting of both transitions from the
dynamics of the system. The doublet formed by the inner peaks is
therefore strong and clearly identifiable in an experiment. On the
other hand, in the direct exciton emission, the counterparts of
Eqs.~(\ref{eq:WedNov26131557GMT2008}) are manifold-independent and
equal for both the inner and outer peaks:
\begin{subequations}
  \label{eq:ThuNov27133856GMT2008}
  \begin{align}
    \label{eq:ThuNov27133905GMT2008}
    &\mathrm{I}_{\sigma}^{(\pm\rightarrow\mp)}\propto|\bra{n-1,\mp}\sigma\ket{n,\pm}|^2=1/4\,,\\
    &\mathrm{I}_{\sigma}^{(\pm\rightarrow\pm)}\propto|\bra{n-1,\pm}\sigma\ket{n,\pm}|^2=1/4\,.
  \end{align}
\end{subequations}
In this case, therefore, one can expect similar strength of
transitions for both the inner and outer peaks with a richer multiplet
structure for the direct exciton emission.

In Fig.~\ref{fig:SunOct19212128BST2008}, we give an overview of the PL
spectra as $P_\sigma$ is varied from very small to very large
values. For point~1, as we already noted, the cavity pumping plays a
relatively minor quantitative role. Therefore we only show two cases,
of no-cavity pumping (first row) and high-cavity pumping (second
row). As can be seen, there is no strong difference from one spectra
with no cavity pumping to its counterpart with large cavity
pumping. Third row shows the direct exciton emission that, with no
cavity pumping, corresponds to the first row. Indeed, one can observe
the richer multiplet structure up to $P_\sigma\approx 0.5g$ in the
direct exciton emission, whereas only inner peaks are neatly manifest
in the cavity emission.  This region corresponds to a quantum regime
with a few quanta of excitations (and subpoissonian particle number
distribution, $g^{(2)}<1$) giving rise to clearly resolvable peaks,
attributable to the Hamiltonian manifolds. Therefore, a good system
(high~$Q$ and $g$) and a good QD (two-level) emitter suffice to easily
and clearly observe quantum effects. There is no need of pumping
harder than it has been done in present systems so far.

In the region $g<P_\sigma<30g$, the photon fluctuations are those of a
coherent, classical state, $g^{(2)}=1$. Increasing pumping with the
intention to penetrate further into the nonlinearity, merely collapses
the multiplet structure into a single line, as far as cavity emission
is concerned. However, this does not mean that the system is in weak
coupling. In the direct exciton emission, the rich SC fine structure
has turned into a \emph{Mollow triplet}~\cite{mollow69a}, that we
discuss in depth below. In this region, the first manifolds have
crossed to WC but higher manifolds retain SC, bringing the system into
lasing. At this point, a change of realm should be performed favoring
a classical description. A last transition into thermal light and WC,
due to saturation and self-quenching, takes place at
$P_\sigma\approx30g$ that leads to a single central peak in the
spectra.

\begin{figure}[hbpt]
  \centering
  \includegraphics[width=\linewidth]{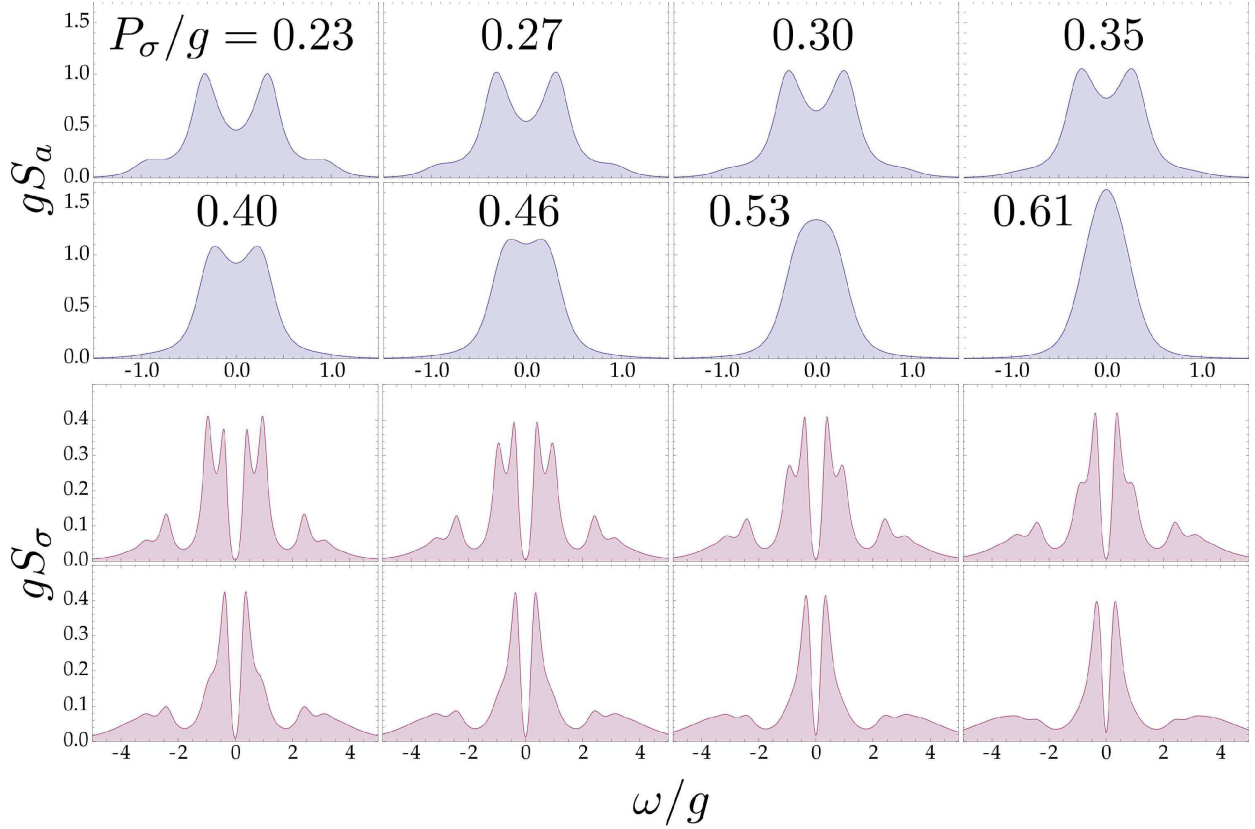}
  \caption{(Color online). Point~1 of
    Fig.~\ref{TueSep16163717BST2008}.  Details of the loss of the
    multiplet structure with increasing exciton pumping and zero
    cavity pumping. The two upper rows (blue) correspond to the cavity
    emission~$S_a(\omega)$ and the two lower (violet) to the exciton
    direct emission~$S_\sigma(\omega)$. The spectral structure is
    richer in the exciton spectra that develops a Mollow triplet-like
    emission.}
  \label{fig:SunOct19212500BST2008}
\end{figure}

\begin{figure}[hbpt]
  \centering
  \includegraphics[width=\linewidth]{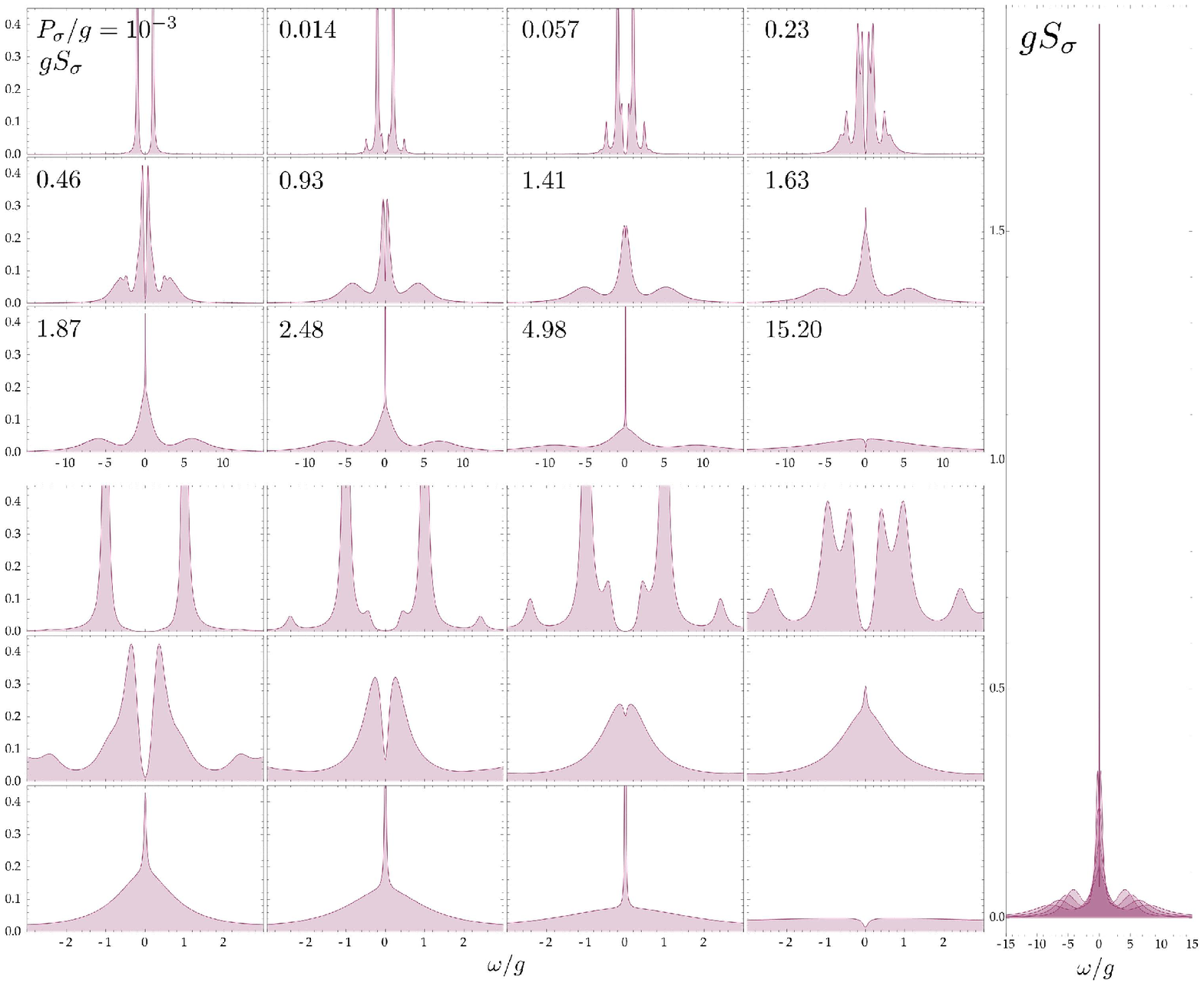}
  \caption{(Color online). Point~1 of
    Fig.~\ref{TueSep16163717BST2008}.  Incoherent Mollow triplets are
    observed in the exciton direct emission with broad satellite peaks
    at approximately~$\pm2\sqrt{n_a}$ and a strong narrow central peak
    taking over a narrow resonance. Three upper rows show the spectra
    over the interval~$|\omega|\le15g$ allowing to see the
    satellites. Three lower rows are the same in the
    window~$|\omega|\le3g$, allowing to see the narrow resonance and
    peak that sit at the origin. Values of the electronic pumping are
    given in the frame of the first three rows. Cavity pumping is zero
    but influences very little the Mollow triplets.  Rightmost figure
    superposes various spectra at increasing electronic pumping,
    showing the drift and broadening of the satellites, and putting to
    scale the very strong coherent feature at the origin. The
    incoherent Mollow triplet appears thus very differently from its
    counterpart under coherent excitation.}
  \label{fig:SunOct19212500BST20082}
\end{figure}

In Fig.~\ref{fig:SunOct19212500BST2008}, we take a closer look into
Fig.~\ref{fig:SunOct19212128BST2008} in the region of the loss of the
doublet of inner peaks with increasing electronic pumping, where the
system starts to cross from the quantum to the classical regime.  In
the cavity emission, the doublet of inner peaks collapses into a
single line that is going to narrow as the system lases. At the same
time, a strikingly richer structure and regime transition is observed
in the direct exciton emission. As the peaks are more clearly resolved
as explained before [cf.~Eqs.~(\ref{eq:WedNov26131557GMT2008})
and~(\ref{eq:ThuNov27133856GMT2008})], the ``melting'' of the
Jaynes-Cummings ladder into a classical structure is better tracked
down. Indeed, as pumping is increased, broadening of these lines
starts to unite them together into an emerging structure of a much
less reduced complexity, namely a triplet.  In
Fig.~\ref{fig:SunOct19212500BST20082}, we provide another zoom of the
overall picture given by Fig.~\ref{fig:SunOct19212128BST2008}, this
time for the direct exciton emission exclusively. First three rows
show the evolution with electronic pumping~$P_\sigma$ (values in
inset) over a wide range of frequencies, up to~$\pm15g$, while the
three last rows show the very same spectra, with a one-to-one mapping
with previous rows, only in the range of frequencies~$\pm3g$.  The
transition manifests to different scales, with a rich fine multiplet
structure in the quantum regime, as seen in the zoomed-in region, to a
monolithic triplet at higher pumpings, as seen in the enlarged
region. On the right, spectra are superimposed to follow their
evolution with pumping. The two satellites peaks, at
approximately~$\pm2\sqrt{n_a}$, drift apart from the main central one
with increasing excitation, and in this sense behave as expected from
a Mollow triplet. Various deviations are however observed, of a more
or less striking character. The most astonishing feature is the
emergence of a very sharp and narrow peak in the center, that has been
plotted with its total intensity on the right panel to give a sense of
its magnitude. It is clearly seen in the zoomed-region how this peak
arises on top of the broad mountain of inner peaks, surviving the
collapse of the fine structure in the classical regime. This thin
central resonance appears when a large truncation is needed. It is a
sum of many contributing peaks centered at zero, most of them with
very small intensities. This region therefore shows all the signs of a
transition from a quantum to a classical system. At low pumping, the
inner peaks of all quadruplets coming from low order manifolds are
placed approximately at $\pm(\sqrt{n}-\sqrt{n-1})\neq0$. Even when
they are summed up to produce the total spectrum, the nonlinear
doublet is still resolved. At around $P_\sigma \approx 1.5g$,
manifolds high enough are excited so that for them
$\pm(\sqrt{n}-\sqrt{n-1})\approx 0$. This is a feature of a classical
field resulting in a Mollow triplet.  Note that nothing of this sort
is observed in the cavity emission. The Mollow triplet, whether in
atomic physics with coherent excitation or in semiconductor physics
with incoherent pumping, is a feature of the quantum emitter itself,
when it is directly probed. There is therefore a strong motivation
here to detect leak emission of semiconductor structures.  The overall
features of this ``incoherent Mollow triplet'' differ from its
counterpart namesake in the strong asymmetry of the satellites and
their increased broadenings with pumping.

\begin{turnpage}
\begin{figure}[hbpt]
  \centering
  \includegraphics[width=\linewidth]{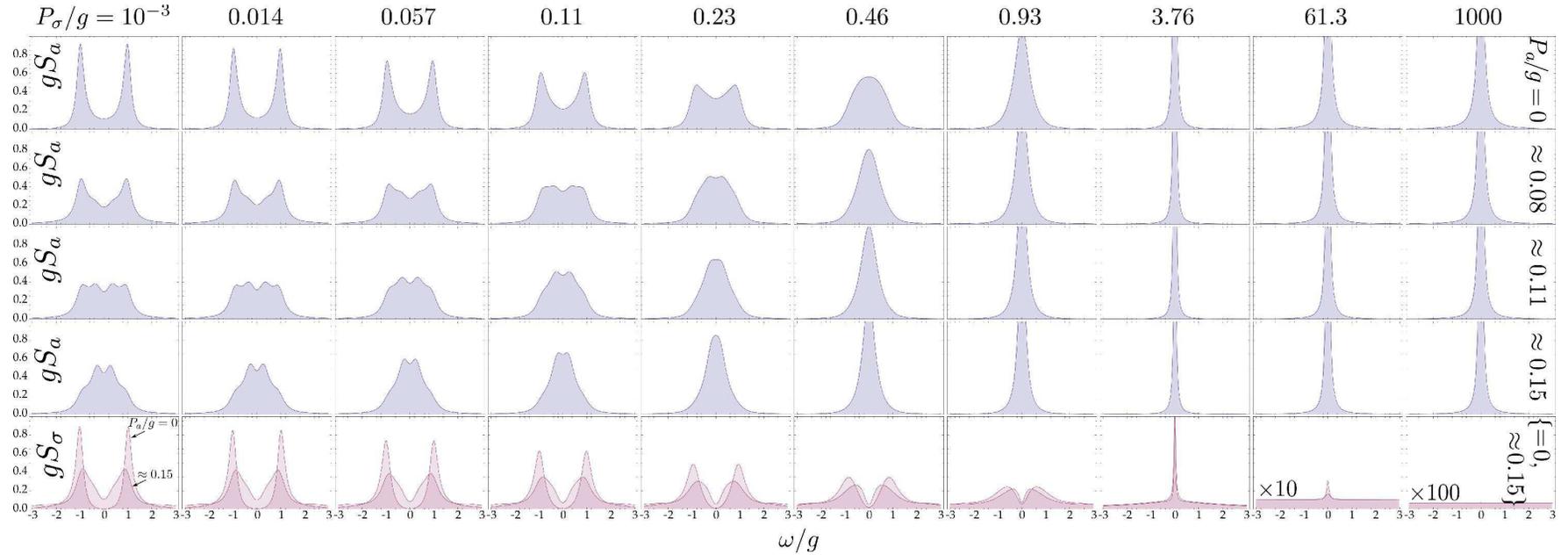}
  \caption{(Color online).  Same as
    Fig.~\ref{fig:SunOct19212128BST2008} over the same range of
    $P_\sigma$, but for Point 2 of
    Fig.~\ref{TueSep16163717BST2008}. In this case, cavity pumping has
    a strong influence on the cavity luminescence spectra, so we show
    more cases, namely $P_a/g=0$ (upper row), $\approx 0.08$ (second),
    $\approx 0.11$ (third) and $\approx0.15$ (fourth) as well as the
    exciton direct emission spectra~$S_\sigma$ in the fifth columns
    with two cases of cavity pumping, $P_a/g=0$ (outer peaks)
    and~$\approx0.15$, corresponding to first and fourth rows of the
    cavity emission.  Exciton spectra are less qualitatively affected
    by the cavity pumping. With electronic pumping only, no particular
    feature is observed in the cavity emission. In this case, cavity
    pumping makes a huge difference by revealing the underlying
    Jaynes-Cummings ladder.}
  \label{fig:TueOct21193007BST2008}
\end{figure}
\end{turnpage}

In Fig.~\ref{fig:TueOct21193007BST2008}, we show for Point 2 a similar
overall picture as Fig.~\ref{fig:SunOct19212128BST2008} does for
Point~1. Point 2 has larger dissipation and to current estimates,
corresponds more closely to the best systems available at the time of
writing. As opposed to Point 1, a small cavity pumping has a strong
influence on the result, so we display more cases, namely those that
range from no cavity pumping (first row) to large cavity pumping
($P_a=g/5$, 4th row) with two intermediate cases showing the transfer
of the emission from the linear Rabi doublet to the inner peaks
arising from transitions between higher manifolds.

\begin{figure}[hbpt]
  \centering
  \includegraphics[width=\linewidth]{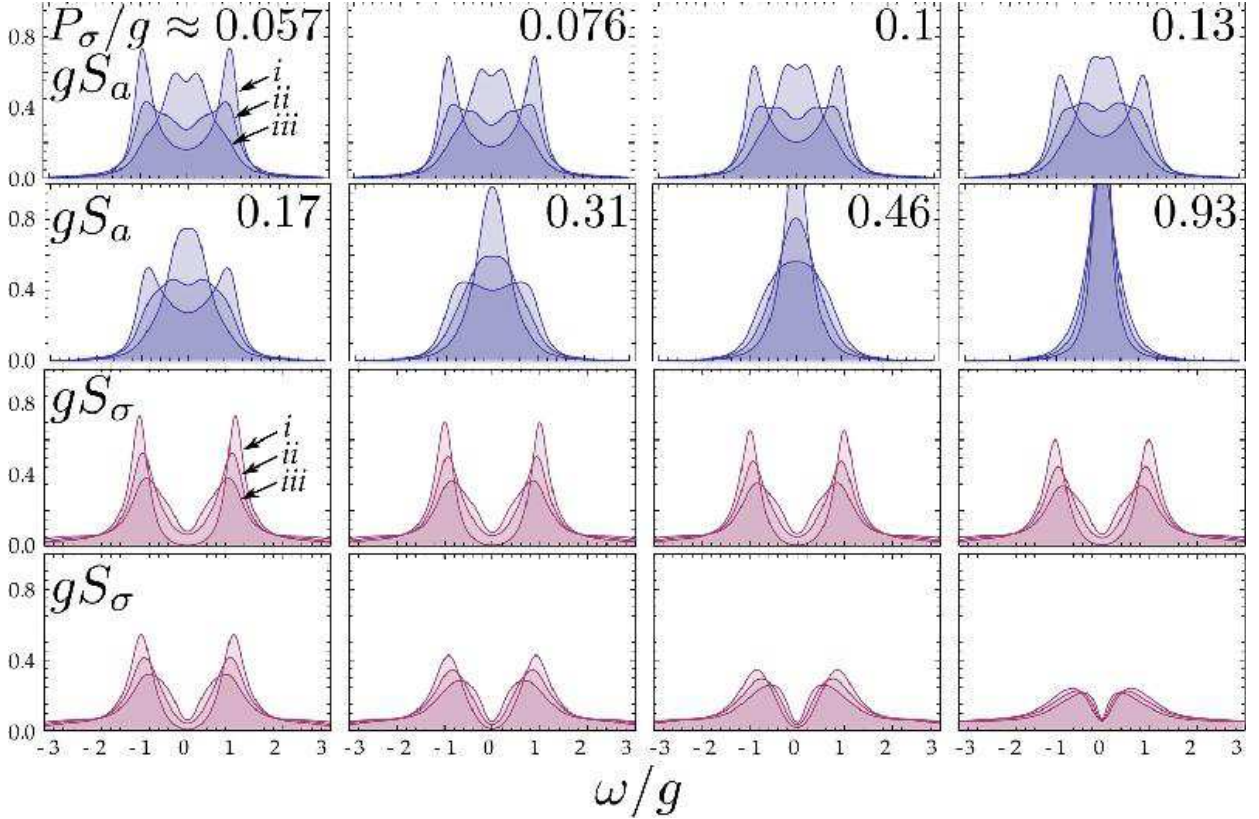}
  \caption{(Color online). Point~2 of
    Fig.~\ref{TueSep16163717BST2008}.  Details of the loss of the
    multiplet structure with increasing exciton pumping.  Two upper
    rows (blue) correspond to the cavity emission~$S_a(\omega)$ and
    two lower (violet) to the corresponding exciton direct
    emission~$S_\sigma(\omega)$ for $P_a/g=0$ ($i$), $\approx0.076$
    ($ii$) and $\approx0.15$ ($iii$) (higher pumping corresponds to
    innermost peaks). Cavity pumping is essential in such a system to
    reveal the Fermionic nature of the QD emitter.}
  \label{fig:WedOct22211718BST2008}
\end{figure}

\begin{figure}[hbpt]
  \centering
  \includegraphics[width=.75\linewidth]{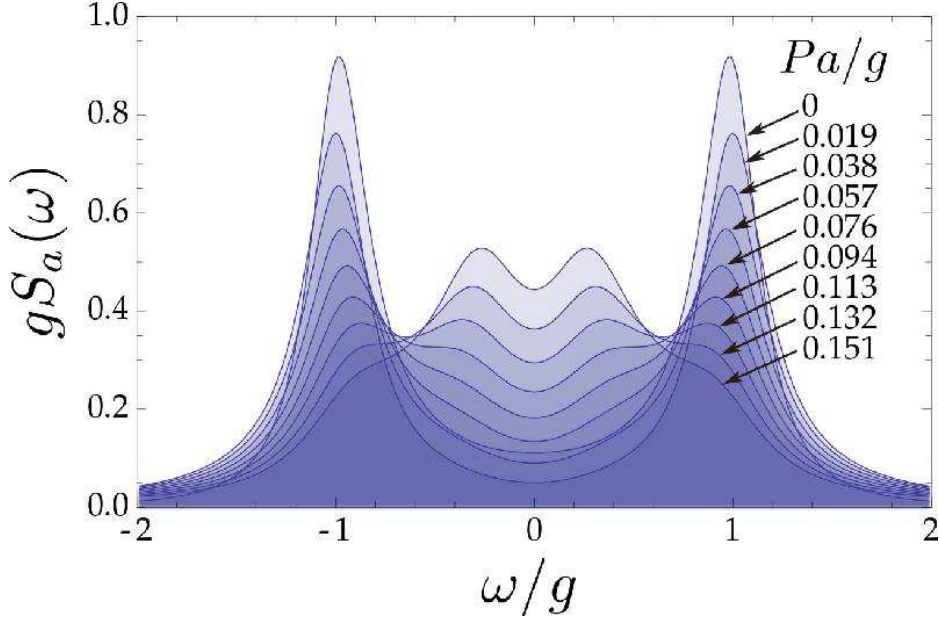}
  \caption{(Color online). Detail of $S_a(\omega)$ for Point~2 of
    Fig.~\ref{TueSep16163717BST2008} at vanishing $P_\sigma$ for the
    values of $P_a/g$ indicated (higher pumpings correspond to
    innermost peaks). In a reasonably good QD--cavity system, strong
    deviations from the linear regime are observed in the emission
    spectrum, revealing the Jaynes-Cummings fork. The quantum features
    are made more obvious by increasing the cavity pumping, with a
    neat renormalization of the dominant doublet even if the
    quadruplet cannot be resolved experimentally.}
  \label{fig:SunOct19212500BST20083}
\end{figure}

The fifth row shows the corresponding direct exciton emission, for the
extreme cases of no (1st row) and highest (4th) cavity pumping.  The
cavity pumping has the important role of revealing the quantum
nonlinearity of the system, that was obvious for reference point~1 in
any case but is now invisible in the first row, where at increasing
electronic pumping, the vacuum Rabi doublet undergoes a rather dull
collapse.  The same spectra could be expected from a linear (bosonic)
model, in the appropriate range of parameters. This is particularly
evident in Fig.~\ref{fig:WedOct22211718BST2008} where three cases of
cavity pumping (none, intermediate, and large) are shown for various
electronic pumping, both for the cavity and direct emission. Outer
lines correspond to zero and inner lines to larger cavity
pumping. Note how the intermediate cavity pumping cases display
obvious deviation from a bosonic model, that has essentially the shape
of a doublet of Lorentzian peaks (with a dispersive correction that
has little bearing on the qualitative aspect of the final
result). Cavity pumping literally unravels the nonlinearity. The case
of intermediate pumping is the most determining in this aspect as far
as cavity emission is concerned, while higher cavity pumpings are more
favorable for uncovering quantum features from the direct exciton
emission. This is mainly for two reasons. One has to do with the
influence of what effective quantum state is realized in the system,
that we will discuss in more details in connection with the third
reference point. The other being the excitation of higher manifolds
from the Jaynes-Cummings ladder, that are now less accessible because
of the larger dissipation rates. Note how the disappearance of the
vacuum Rabi doublet with increasing $P_\sigma$ (with no cavity
pumping), is of a different character than for Point 1, where higher
$P_\sigma$ resulted in an excitation of the upper manifolds and a
transfer of the dynamics higher in the Jaynes-Cummings ladder, whereas
in this case it essentially results in a competition between only the
first and second manifold transitions. Cavity pumping can help
climbing the ladder with no prejudice to broadening. Finally, even if
blurry resolution or statistical noise of an actual experiment would
cast doubt on the presence of a quadruplet in such a structure, the
transfer with increasing cavity pumping of the emission from outer
(linear Rabi) to inner peaks (from the second manifold transitions in
this case) makes it clear that the underlying statistics is of a Fermi
rather than of a Bose character. In
Fig.~\ref{fig:SunOct19212500BST20083}, we show the
case~$P_\sigma=10^{-3}g$ for such an increasing cavity pumping for a
detailed appreciation of the previous statement. A very close look
might still suggest that the case~$P_a=0$ (outer peaks) still has a
small deviation from the linear model that would betray, in a very
finely resolved experiment, its non-bosonic or nonlinear
character. Counter to intuition, this is better seen for vanishing
electronic pumping, as otherwise the lines are broadened according to
Eq.~(\ref{eq:MonAug11125316CEST2008}) and this dampens the inner
nonlinear peaks. Note, on the other hand, how cavity pumping
unambiguously settles the issue.

\begin{figure}[hbpt]
  \centering
  \includegraphics[width=\linewidth]{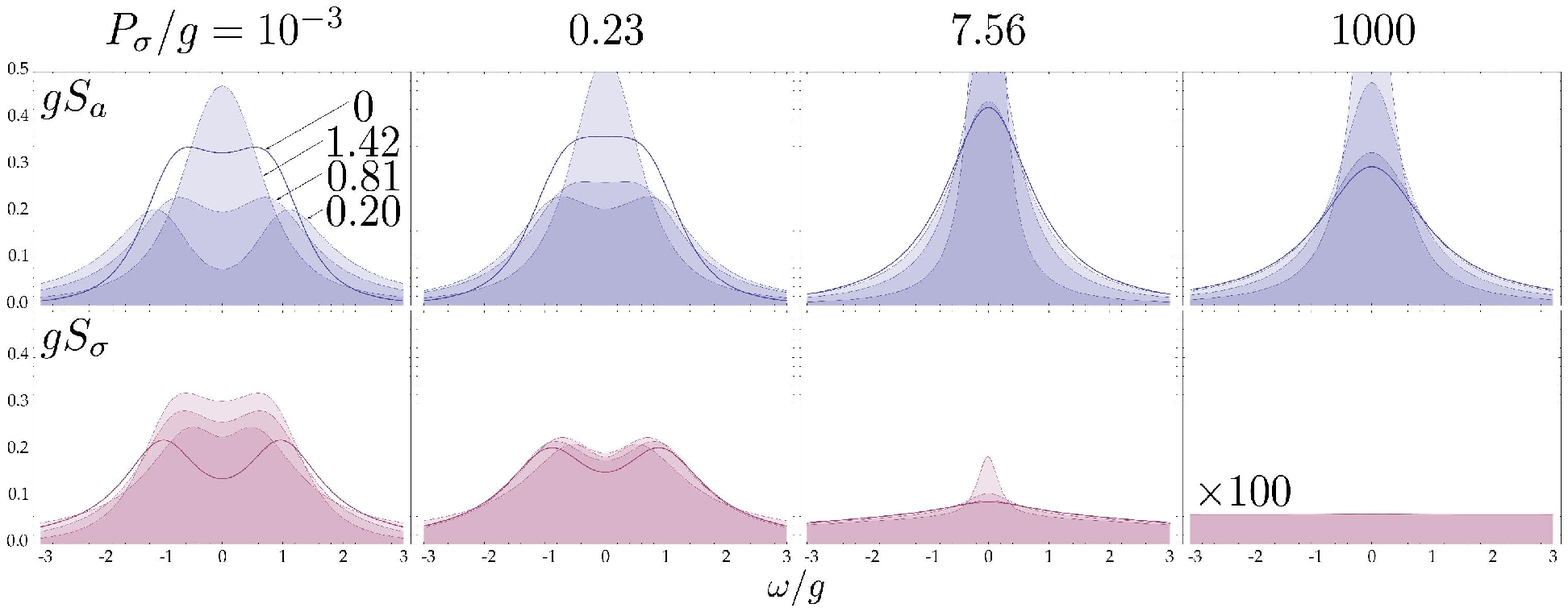}
  \caption{(Color online). Point 3 of
    Fig.~\ref{TueSep16163717BST2008}. Spectral emission for the
    indicated electronic pumping~$P_\sigma/g$: $10^{-3}$ (1st column),
    $\approx0.23$ (2nd), $\approx7.56$ (3rd, lasing) and $1000$ (4th,
    quenching), for $P_a/g=0$ (thick line with no coloring),
    $\approx0.20$, $\approx0.81$ and~$\approx1.42$ as indicated in the
    top left panel, and similarly for others (apart from the case
    $P_a=0$, inner peaks corresponds to higher pumpings). In this
    system, broadening is always too high to allow any manifestation
    of the underlying Jaynes-Cummings structure. The structure could
    be mistaken for a bosonic system (or the other way
    around). Especially, cavity pumping helps observation of the Rabi
    doublet in the same way as for bosons
    modes~\cite{elena_laussy08a}.}
  \label{fig:WedOct29103030GMT2008}
\end{figure}

Finally, we turn to point~3 of Fig.~\ref{TueSep16163717BST2008}, i.e.,
to the case with high dissipation rates. In this case, as shown in
Fig.~\ref{fig:WedOct29103030GMT2008}, the Jaynes-Cummings structure is
not probed and the spectra are mere Rabi doublets, closing in the
WC. These features, by themselves, without a quantitative comparison
with other models, say nothing about the nature of the emitter. The
linear model, always in SC for this system, leads to a well resolved
Rabi doublet in both channels of emission at all pumpings. In the
present model, too high a cavity pumping brings the first manifold
into WC. In contrast, a small cavity pumping again helps to resolve
the Rabi doublet. The main physics at work here is the one that has
been amply detailed in part~1 of this study, in the linear case,
namely, the effective quantum state realized in the system by the
interplay of pumpings and decay. A photon-like quantum state has
dispersive corrections that push apart the dressed states
(Lorentzians) and therefore enhances the visibility and splitting of
the lines. The argument does not adapt itself exactly, for instance
line-splitting is not helped or revealed in the exciton emission for
most cases, because in this case exciton broadening always spoils
resolution of the splitting. In the linear case, there would be a
complete symmetry (enhanced splitting in one channel of detection
implies reduced visibility in the other channel).  This shows again
that the same system would lead, in the nonlinear regime, to different
results with a bosonic or a fermionic exciton, although both spectra
can only feature a doublet or a singlet. A fundamental difference
between the models is that the pumps $P_a$, $P_b$, always reduce the
total broadening of the lines ($\Gamma_+$) while $P_\sigma$ increases
it. The contribution of pump to the line positions differs greatly
from the bosons, as not only $P_\sigma$ carries a different sign but
also this contribution depends on the manifold. The statistics make
also an important difference. Opposite to the wide variety of photon
distributions found with a fermion model, cavity and exciton are
always in a thermal state for a boson, without quenching or lasing.
The issue of the underlying statistics could therefore be settled in
photon-counting experiment. Fig.~\ref{fig:SunOct5183310BST2008} shows
that such systems (especially when~$\gamma_a\gg1$
and~$\gamma_\sigma\rightarrow0$) have the advantage over better
cavities that at low electronic pumping and \emph{vanishing} cavity
pumping, the system generates antibunched light, suitable for
single-photon emitters (though not on demand).

\section{Luminescence spectra with detuning}
\label{SunNov16140046GMT2008}

In semiconductors, the detuning between bare modes is a parameter that
can easily be varied and which provides useful information of the SC
physics. Strong coupling is better studied at resonance, and detuning
is mainly used to help locate it, by finding the point where
anticrossing is maximum and level repulsion stationary. In a fitting
analysis of an experiment, it brings a lot of additional data at the
cost of only one additional fitting parameter. In the Fermion case, it
also has the benefit of uncovering new qualitative behavior of the PL
lineshapes, that are strongly restricted by symmetry at resonance.

\begin{figure}[hbpt]
  \centering
  \includegraphics[width=\linewidth]{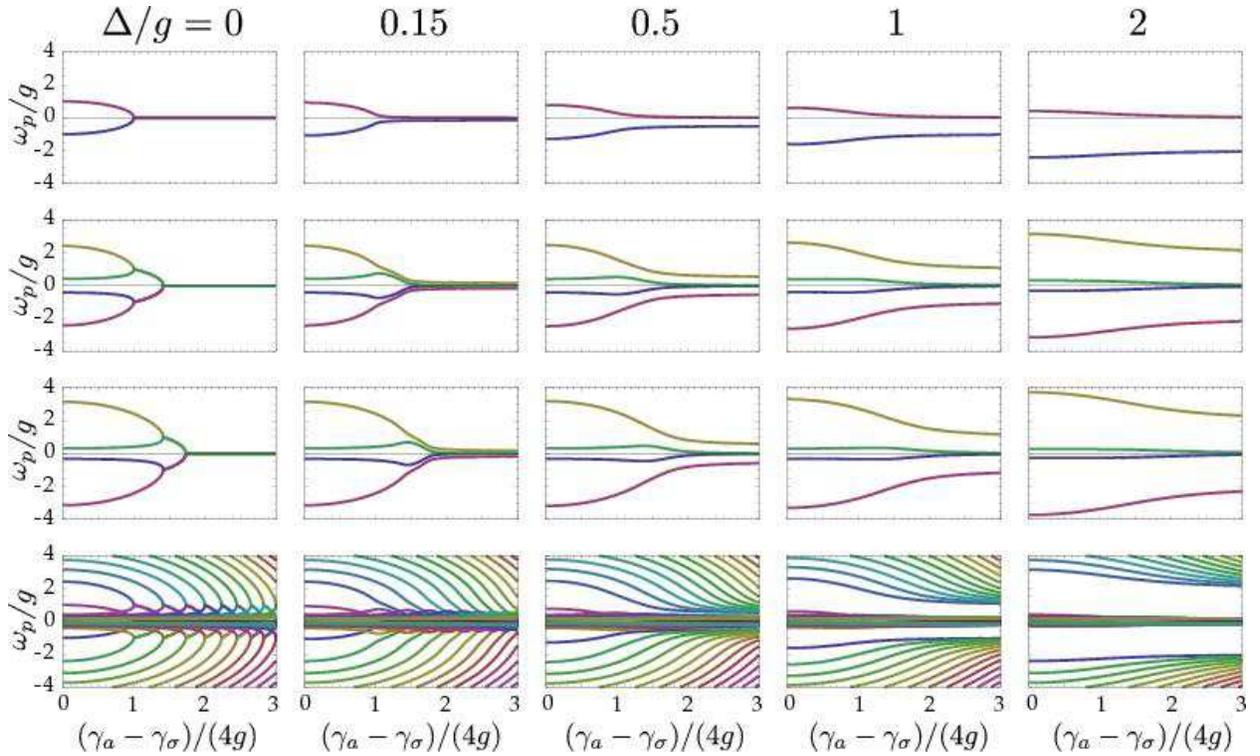}
  \caption{(Color online). Positions $\omega_p/g$ of the lines around
    $\omega_a=0$ in the luminescence spectrum with detuning and in the
    absence of pump. Columns correspond to various detunings, first
    column being the case of resonance
    (cf.~Fig.~\ref{fig:TueAug12144011CEST2008}). First three rows show
    in isolation the first, second and third manifold,
    respectively. First manifold corresponds to the Boson or linear
    case~\cite{elena_laussy08a}. Fourth row shows all manifolds
    together. Left-bottom panel is detailed for positive~$\omega_p$ in
    Fig.~\ref{fig:TueAug12144011CEST2008}.}
  \label{fig:ThuOct30153849GMT2008}
\end{figure}

Fig.~\ref{fig:ThuOct30153849GMT2008} shows the vanishing pumping case
of~$\omega_p$ in Eq.~(\ref{eq:SatAug2111206CEST2008}) with detuning,
i.e., the imaginary part of Eq.~(\ref{eq:TueAug12190002CEST2008}) for
the first row that corresponds to the first manifold (also, the boson
case) and of Eq.~(\ref{eq:TueAug12185731CEST2008}) for the second and
third rows, that corresponds to the second and third manifold,
respectively. Fourth row is a superposition of all manifolds up to
the 15th one. Detuning is varied in columns, from no detuning (first
column) to twice the coupling strength (fifth column). Negative
detunings are symmetric with respect to the $x$~axis.

\begin{figure}[hbpt]
  \centering
  \includegraphics[width=.8\linewidth]{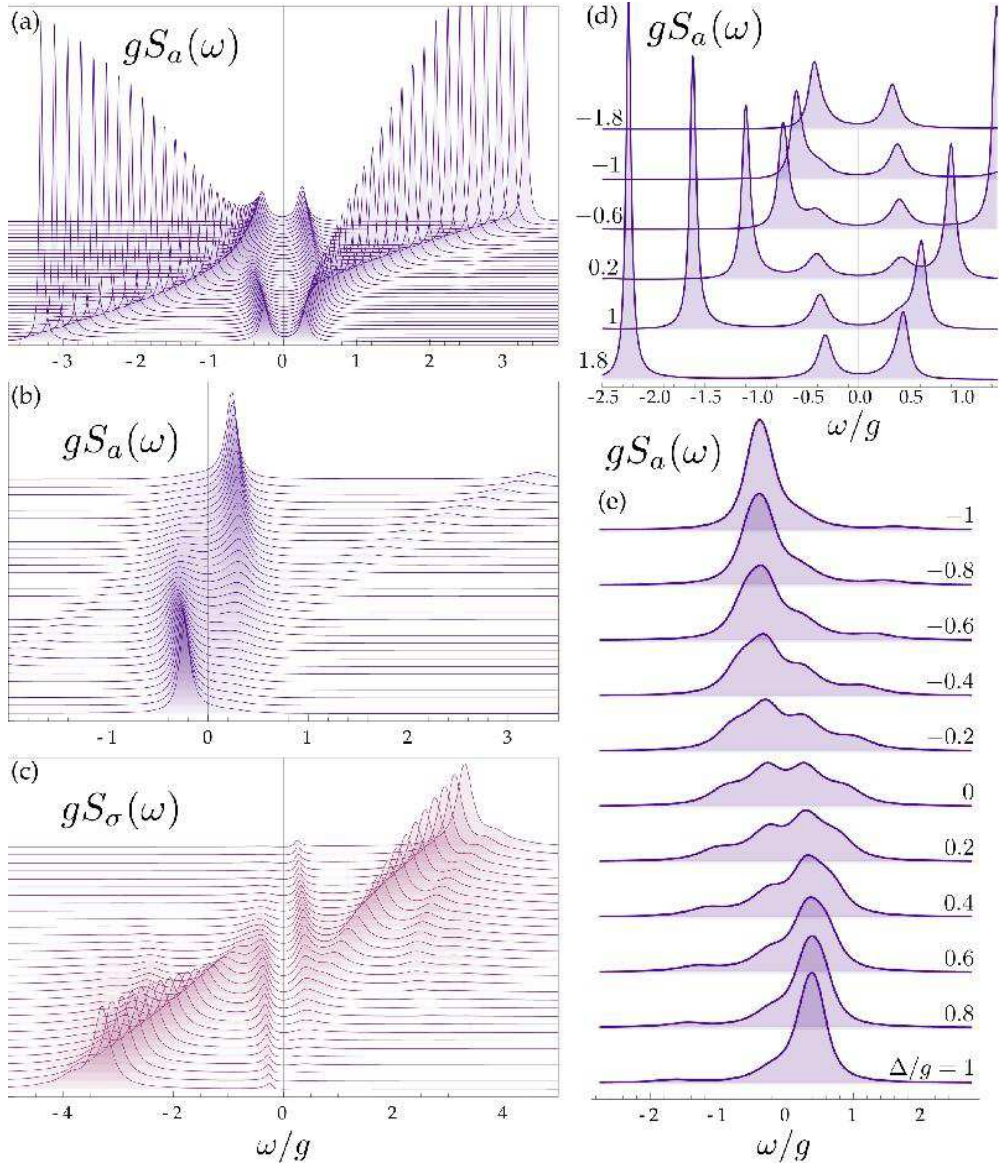}
  \caption{(Color online). Anticrossing of the luminescence lines as
    detuning~$\Delta=\omega_a-\omega_\sigma$ is varied. Here,
    $\omega_a=0$ is fixed and the QD bare energy is tuned from below
    the cavity (positive detuning) to above (negative
    detuning). Panels (a)-(d) correspond to Point 1 and panel (e) to
    Point 2. (a)-(d) are at zero cavity pumping, $P_a=0$. (a) and (d)
    are for $P_\sigma=0.03g$ [(d) is a zoom of (a)] and (b)-(c) for
    $P_\sigma=0.3g$. (a), (b), (d) are the cavity
    emission~$S_a(\omega)$ and~(c) the direct exciton
    emission~$S_\sigma(\omega)$. (e) is for $P_\sigma=10^{-3}g$ and
    $P_a=g/5$ (cf., 7th row, 1st column of
    Fig.~\ref{fig:TueOct21193007BST2008}). The nonlinear central peaks
    give rise to very characteristic anticrossing profiles.}
  \label{fig:ThuOct30102109GMT2008}
\end{figure}

The line opening is common to all manifolds, but note the different
behavior of the first manifold (linear or boson case) and higher
manifolds: in the first case, one line collapses towards the center
(on the cavity mode) while the other recedes away, towards the exciton
mode. In the nonlinear case, there is up to four lines, and outer
lines are both repelled away while inner lines get both attracted
towards the cavity mode, at the center.  As we discussed, the total
doublet of inner peaks is intense and will dominate. For cases with
high dissipation, there is little or no particular insights to be
gained from detuning, as, again, most features are lost in
broadening. We restrict out attention to Points~1 and~2 in what
follows. In Fig.~\ref{fig:ThuOct30102109GMT2008}, PL with detuning are
shown for Point~1 in panels (a)--(d) and for Point~2 in
panel~(e). Panel~(d) is a magnified view of panel~(a). It is seen
clearly how the doublet of inner peaks essentially remains fixed at
its resonance position independently of the exciton position. Only at
very high detunings does the doublet collapse onto the center. The
vacuum Rabi doublet however appears as an anticrossing of the exciton
bare mode with the doublet of inner peaks (that eventually becomes the
cavity bare mode). Panel~(a) is at small electronic pumping and~(b),
(c) at ten time larger electronic pumping (both no cavity pumping),
for the cavity and direct exciton emission, respectively. Again, lower
electronic pumping is more prone to reveal rich quantum features. In
panel~(b) only the inner nonlinear doublet is visible, with a transfer
of the emission intensity from one peak (essentially fixed) to the
other. The resonance case is plotted in the third panel on first row
(third row for the exciton emission) of
Fig.~\ref{fig:SunOct19212500BST2008}. The linear Rabi doublet, which
trace is seen faintly undergoing anticrossing with the pinned central
peaks, provides small shoulders.  In general, PL with detuning in the
Fermi case shows a very characteristic behavior, that cannot be
mistaken with a conventional (bosonic) anticrossing experiment.

In panel~(e), the case of a more realistic system is shown with
detuning. The pinning of the inner peaks is less obvious in this case,
although if one draws a vertical line at the resonance, through the
minimum of the doublet, one observes that this minimum is fixed. As a
result, triplets are obtained in the cavity emission spectra, that are
of a very distinct nature than the Mollow triplet observed in the side
(exciton) emission of Point~1. The triplets involving the nonlinear
doublet are a manifestation of the quantum regime overcoming
broadening while the Mollow triplet is a manifestation of the lasing
regime.

\section{Discussion and Conclusions}
\label{FriNov7121705GMT2008}

We now give an overview of our results on the spectral lineshapes of a
QD that obeys Fermi statistics, in SC with the single mode of a
semiconductor microcavity, and extend our previous exposition with a
more general discussion. The main results of our analysis are as
follow.

Manifestations of nonlinearities in the SC physics of a genuine
quantum emitter are, counter to naive expectations, no better sought
at high pumpings, looking forward to large number of excitations. The
quantum regime involves a few quanta only. It is achieved and better
manifests with \emph{low pumpings} in \emph{high quality samples}
(meaning that the exciton-photon coupling should be much higher than
the decays and dissipation rates). Higher pumpings will cross to the
classical regime where the quantumness of the system---that manifests
with clearly separated peaks attributable to well identified
transitions in the Jaynes-Cummings ladder---will give over to a very
large number of very small correlators, conveying that a continuous
field is taking over quantum discretization. High pumping that bring
the system into a lasing regime has a behavior that is better
expressed by classical fields and classical physics. The crossover
from a quantum to a classical description is an interesting problem
that such a model---the dissipative Jaynes-Cummings Hamiltonian with
incoherent pumping--seems particularly well suited to track
theoretically.

Cavity pumping is an important factor to take into account. First
because of its relevance in an actual experiment, where it can arise
due to secondary effects such as other dots (not in SC) emitting in
the cavity, temperature, or a variety of other factors. It could
conceivably also be input directly by the experimentalist. Cavity
pumping has many virtues for the physics of SC in a
semiconductor. Because the typical type of excitation is electronic
and the typical channel of detection is photonic, SC is hampered as
compared to the microwave cavity case where detection and excitation
are on the same footing (both directly on the atom).  A cavity
pumping can help balance this situation and provide an effective
photon character to the states realized in the semiconductor,
enhancing or even revealing spectral structures.  This phenomena
manifests also in the Boson case and has been investigated and
explained in its full detailed in part~I of this work. Also in the
Fermion case, cavity pumping is beneficial for the same reasons, and
it can help go beyond the linear regime (with a Rabi doublet) to the
nonlinear quantum regime, typically by making emerge additional
quadruplets of the Jaynes-Cummings, with a doublet of inner peaks to
be sought as the strongest signature.

The counterpart of a Mollow triplet is observed in the best samples,
in the direct exciton emission. It features a narrow resonance in the
center of the spectrum that turns into a sharp emission line. The
incoherent Mollow triplet is a striking manifestation of a crossover
from the quantum to the classical limit, with a series of many peaks
easily identifiable with dressed quantum states, melting into a
monolithic structure of reduced complexity (a triplet) with no
identifiable contributions from separable processes. When the Mollow
triplet is fully formed, the cavity mode is in the lasing regime. The
Mollow triplet is lost as the system is quenched with no return to
quantum behaviors. This provides the general sequence of regimes with
increasing electronic pumping: quantum regime, lasing (classical)
regime and quenched (also classical but thermal) regime.
 
\begin{acknowledgements}
  Most numerical computations presented here have been performed on
  the Iridis cluster of the University of Southampton, for which
  authors gratefully acknowledge the support from Dr.~Ivan Woulton.
  This work has been supported by the Spanish MEC under contracts
  Consolider-Ingenio2010 CSD2006-0019, MAT2005-01388 and
  NAN2004-09109-C04-3 and by CAM under contract S-0505/ESP-0200.  EdV
  acknowledges support of the FPU from the Spanish MEC.
\end{acknowledgements}

\bibliography{Sci,books,lsqdm2,Elena}

\begin{thebibliography}{42}
\expandafter\ifx\csname natexlab\endcsname\relax\def\natexlab#1{#1}\fi
\expandafter\ifx\csname bibnamefont\endcsname\relax
  \def\bibnamefont#1{#1}\fi
\expandafter\ifx\csname bibfnamefont\endcsname\relax
  \def\bibfnamefont#1{#1}\fi
\expandafter\ifx\csname citenamefont\endcsname\relax
  \def\citenamefont#1{#1}\fi
\expandafter\ifx\csname url\endcsname\relax
  \def\url#1{\texttt{#1}}\fi
\expandafter\ifx\csname urlprefix\endcsname\relax\def\urlprefix{URL }\fi
\providecommand{\bibinfo}[2]{#2}
\providecommand{\eprint}[2][]{\url{#2}}

\bibitem[{\citenamefont{Laussy et~al.}(2008{\natexlab{a}})\citenamefont{Laussy,
  del Valle, and Tejedor}}]{elena_laussy08a}
\bibinfo{author}{\bibfnamefont{F.~P.} \bibnamefont{Laussy}},
  \bibinfo{author}{\bibfnamefont{E.}~\bibnamefont{del Valle}},
  \bibnamefont{and} \bibinfo{author}{\bibfnamefont{C.}~\bibnamefont{Tejedor}},
  \bibinfo{journal}{arXiv:0807.3194}  (\bibinfo{year}{2008}{\natexlab{a}}).

\bibitem[{\citenamefont{Jaynes and Cummings}(1963)}]{jaynes63a}
\bibinfo{author}{\bibfnamefont{E.}~\bibnamefont{Jaynes}} \bibnamefont{and}
  \bibinfo{author}{\bibfnamefont{F.}~\bibnamefont{Cummings}},
  \bibinfo{journal}{Proc. IEEE} \textbf{\bibinfo{volume}{51}},
  \bibinfo{pages}{89} (\bibinfo{year}{1963}).

\bibitem[{\citenamefont{Laussy et~al.}(2006)\citenamefont{Laussy, Glazov,
  Kavokin, Whittaker, and Malpuech}}]{laussy06b}
\bibinfo{author}{\bibfnamefont{F.~P.} \bibnamefont{Laussy}},
  \bibinfo{author}{\bibfnamefont{M.~M.} \bibnamefont{Glazov}},
  \bibinfo{author}{\bibfnamefont{A.}~\bibnamefont{Kavokin}},
  \bibinfo{author}{\bibfnamefont{D.~M.} \bibnamefont{Whittaker}},
  \bibnamefont{and} \bibinfo{author}{\bibfnamefont{G.}~\bibnamefont{Malpuech}},
  \bibinfo{journal}{Phys. Rev. B} \textbf{\bibinfo{volume}{73}},
  \bibinfo{pages}{115343} (\bibinfo{year}{2006}).

\bibitem[{\citenamefont{Rudin and Reinecke}(1999)}]{rudin99a}
\bibinfo{author}{\bibfnamefont{S.}~\bibnamefont{Rudin}} \bibnamefont{and}
  \bibinfo{author}{\bibfnamefont{T.~L.} \bibnamefont{Reinecke}},
  \bibinfo{journal}{Phys. Rev. B} \textbf{\bibinfo{volume}{59}},
  \bibinfo{pages}{10227} (\bibinfo{year}{1999}).

\bibitem[{\citenamefont{Zhu et~al.}(1990)\citenamefont{Zhu, Gauthier, Morin,
  Wu, Carmichael, and Mossberg}}]{zhu90a}
\bibinfo{author}{\bibfnamefont{Y.}~\bibnamefont{Zhu}},
  \bibinfo{author}{\bibfnamefont{D.~J.} \bibnamefont{Gauthier}},
  \bibinfo{author}{\bibfnamefont{S.~E.} \bibnamefont{Morin}},
  \bibinfo{author}{\bibfnamefont{Q.}~\bibnamefont{Wu}},
  \bibinfo{author}{\bibfnamefont{H.~J.} \bibnamefont{Carmichael}},
  \bibnamefont{and} \bibinfo{author}{\bibfnamefont{T.~W.}
  \bibnamefont{Mossberg}}, \bibinfo{journal}{Phys. Rev. Lett.}
  \textbf{\bibinfo{volume}{64}}, \bibinfo{pages}{2499} (\bibinfo{year}{1990}).

\bibitem[{\citenamefont{Schneebeli et~al.}(2008)\citenamefont{Schneebeli, Kira,
  and Koch}}]{schneebeli08a}
\bibinfo{author}{\bibfnamefont{L.}~\bibnamefont{Schneebeli}},
  \bibinfo{author}{\bibfnamefont{M.}~\bibnamefont{Kira}}, \bibnamefont{and}
  \bibinfo{author}{\bibfnamefont{S.~W.} \bibnamefont{Koch}},
  \bibinfo{journal}{Phys. Rev. Lett.} \textbf{\bibinfo{volume}{101}},
  \bibinfo{pages}{097401} (\bibinfo{year}{2008}).

\bibitem[{\citenamefont{Steiner et~al.}(2008)\citenamefont{Steiner, Kira, and
  Koch}}]{steiner08a}
\bibinfo{author}{\bibfnamefont{J.~T.} \bibnamefont{Steiner}},
  \bibinfo{author}{\bibfnamefont{M.}~\bibnamefont{Kira}}, \bibnamefont{and}
  \bibinfo{author}{\bibfnamefont{S.~W.} \bibnamefont{Koch}},
  \bibinfo{journal}{Phys. Rev. B} \textbf{\bibinfo{volume}{77}},
  \bibinfo{pages}{165308} (\bibinfo{year}{2008}).

\bibitem[{\citenamefont{Srinivasan and Painter}(2007)}]{srinivasan07a}
\bibinfo{author}{\bibfnamefont{K.}~\bibnamefont{Srinivasan}} \bibnamefont{and}
  \bibinfo{author}{\bibfnamefont{O.}~\bibnamefont{Painter}},
  \bibinfo{journal}{Nature} \textbf{\bibinfo{volume}{450}},
  \bibinfo{pages}{862} (\bibinfo{year}{2007}).

\bibitem[{\citenamefont{Press et~al.}(2007)\citenamefont{Press, G\"otzinger,
  Reitzenstein, Hofmann, L\"offler, Kamp, Forchel, and Yamamoto}}]{press07a}
\bibinfo{author}{\bibfnamefont{D.}~\bibnamefont{Press}},
  \bibinfo{author}{\bibfnamefont{S.}~\bibnamefont{G\"otzinger}},
  \bibinfo{author}{\bibfnamefont{S.}~\bibnamefont{Reitzenstein}},
  \bibinfo{author}{\bibfnamefont{C.}~\bibnamefont{Hofmann}},
  \bibinfo{author}{\bibfnamefont{A.}~\bibnamefont{L\"offler}},
  \bibinfo{author}{\bibfnamefont{M.}~\bibnamefont{Kamp}},
  \bibinfo{author}{\bibfnamefont{A.}~\bibnamefont{Forchel}}, \bibnamefont{and}
  \bibinfo{author}{\bibfnamefont{Y.}~\bibnamefont{Yamamoto}},
  \bibinfo{journal}{Phys. Rev. Lett.} \textbf{\bibinfo{volume}{98}},
  \bibinfo{pages}{117402} (\bibinfo{year}{2007}).

\bibitem[{\citenamefont{Kroner et~al.}(2008)\citenamefont{Kroner, Govorov,
  Remi, Biedermann, Seidl, Badolato, Petroff, Zhang, Barbour, Gerardot
  et~al.}}]{kroner08a}
\bibinfo{author}{\bibfnamefont{M.}~\bibnamefont{Kroner}},
  \bibinfo{author}{\bibfnamefont{A.~O.} \bibnamefont{Govorov}},
  \bibinfo{author}{\bibfnamefont{S.}~\bibnamefont{Remi}},
  \bibinfo{author}{\bibfnamefont{B.}~\bibnamefont{Biedermann}},
  \bibinfo{author}{\bibfnamefont{S.}~\bibnamefont{Seidl}},
  \bibinfo{author}{\bibfnamefont{A.}~\bibnamefont{Badolato}},
  \bibinfo{author}{\bibfnamefont{P.~M.} \bibnamefont{Petroff}},
  \bibinfo{author}{\bibfnamefont{W.}~\bibnamefont{Zhang}},
  \bibinfo{author}{\bibfnamefont{R.}~\bibnamefont{Barbour}},
  \bibinfo{author}{\bibfnamefont{B.~D.} \bibnamefont{Gerardot}},
  \bibnamefont{et~al.}, \bibinfo{journal}{Nature}
  \textbf{\bibinfo{volume}{451}}, \bibinfo{pages}{311} (\bibinfo{year}{2008}).

\bibitem[{\citenamefont{Sanchez-Mondragon
  et~al.}(1983)\citenamefont{Sanchez-Mondragon, Narozhny, and
  Eberly}}]{sanchezmondragon83a}
\bibinfo{author}{\bibfnamefont{J.~J.} \bibnamefont{Sanchez-Mondragon}},
  \bibinfo{author}{\bibfnamefont{N.~B.} \bibnamefont{Narozhny}},
  \bibnamefont{and} \bibinfo{author}{\bibfnamefont{J.~H.}
  \bibnamefont{Eberly}}, \bibinfo{journal}{Phys. Rev. Lett.}
  \textbf{\bibinfo{volume}{51}}, \bibinfo{pages}{550} (\bibinfo{year}{1983}).

\bibitem[{\citenamefont{Inoue et~al.}(2008)\citenamefont{Inoue, Ochiai, and
  Sakoda}}]{inoue08a}
\bibinfo{author}{\bibfnamefont{J.~I.} \bibnamefont{Inoue}},
  \bibinfo{author}{\bibfnamefont{T.}~\bibnamefont{Ochiai}}, \bibnamefont{and}
  \bibinfo{author}{\bibfnamefont{K.}~\bibnamefont{Sakoda}},
  \bibinfo{journal}{Phys. Rev. A} \textbf{\bibinfo{volume}{77}},
  \bibinfo{pages}{015806} (\bibinfo{year}{2008}).

\bibitem[{\citenamefont{Carmichael et~al.}(1989)\citenamefont{Carmichael,
  Brecha, Raizen, Kimble, and Rice}}]{carmichael89a}
\bibinfo{author}{\bibfnamefont{H.~J.} \bibnamefont{Carmichael}},
  \bibinfo{author}{\bibfnamefont{R.~J.} \bibnamefont{Brecha}},
  \bibinfo{author}{\bibfnamefont{M.~G.} \bibnamefont{Raizen}},
  \bibinfo{author}{\bibfnamefont{H.~J.} \bibnamefont{Kimble}},
  \bibnamefont{and} \bibinfo{author}{\bibfnamefont{P.~R.} \bibnamefont{Rice}},
  \bibinfo{journal}{Phys. Rev. A} \textbf{\bibinfo{volume}{40}},
  \bibinfo{pages}{5516} (\bibinfo{year}{1989}).

\bibitem[{\citenamefont{Andreani et~al.}(1999)\citenamefont{Andreani,
  Panzarini, and G\'erard}}]{andreani99a}
\bibinfo{author}{\bibfnamefont{L.~C.} \bibnamefont{Andreani}},
  \bibinfo{author}{\bibfnamefont{G.}~\bibnamefont{Panzarini}},
  \bibnamefont{and} \bibinfo{author}{\bibfnamefont{J.-M.}
  \bibnamefont{G\'erard}}, \bibinfo{journal}{Phys. Rev. B}
  \textbf{\bibinfo{volume}{60}}, \bibinfo{pages}{13276} (\bibinfo{year}{1999}).

\bibitem[{\citenamefont{Cui and Raymer}(2006)}]{cui06a}
\bibinfo{author}{\bibfnamefont{G.}~\bibnamefont{Cui}} \bibnamefont{and}
  \bibinfo{author}{\bibfnamefont{M.~G.} \bibnamefont{Raymer}},
  \bibinfo{journal}{Phys. Rev. A} \textbf{\bibinfo{volume}{73}},
  \bibinfo{pages}{053807} (\bibinfo{year}{2006}).

\bibitem[{\citenamefont{Auff\`eves et~al.}(2008)\citenamefont{Auff\`eves,
  Besga, G\'erard, and Poizat}}]{auffeves08a}
\bibinfo{author}{\bibfnamefont{A.}~\bibnamefont{Auff\`eves}},
  \bibinfo{author}{\bibfnamefont{B.}~\bibnamefont{Besga}},
  \bibinfo{author}{\bibfnamefont{J.-M.} \bibnamefont{G\'erard}},
  \bibnamefont{and} \bibinfo{author}{\bibfnamefont{J.-P.}
  \bibnamefont{Poizat}}, \bibinfo{journal}{Phys. Rev. A}
  \textbf{\bibinfo{volume}{77}}, \bibinfo{pages}{063833}
  (\bibinfo{year}{2008}).

\bibitem[{\citenamefont{Mollow}(1969)}]{mollow69a}
\bibinfo{author}{\bibfnamefont{B.~R.} \bibnamefont{Mollow}},
  \bibinfo{journal}{Phys. Rev.} \textbf{\bibinfo{volume}{188}},
  \bibinfo{pages}{1969} (\bibinfo{year}{1969}).

\bibitem[{\citenamefont{Savage}(1989)}]{savage89a}
\bibinfo{author}{\bibfnamefont{C.~M.} \bibnamefont{Savage}},
  \bibinfo{journal}{Phys. Rev. Lett.} \textbf{\bibinfo{volume}{63}},
  \bibinfo{pages}{1376} (\bibinfo{year}{1989}).

\bibitem[{\citenamefont{Freedhoff and Quang}(1994)}]{freedhoff94a}
\bibinfo{author}{\bibfnamefont{H.}~\bibnamefont{Freedhoff}} \bibnamefont{and}
  \bibinfo{author}{\bibfnamefont{T.}~\bibnamefont{Quang}},
  \bibinfo{journal}{Phys. Rev. Lett.} \textbf{\bibinfo{volume}{72}},
  \bibinfo{pages}{474} (\bibinfo{year}{1994}).

\bibitem[{\citenamefont{Barchielli and Pero}(2002)}]{barchielli02a}
\bibinfo{author}{\bibfnamefont{A.}~\bibnamefont{Barchielli}} \bibnamefont{and}
  \bibinfo{author}{\bibfnamefont{N.}~\bibnamefont{Pero}}, \bibinfo{journal}{J.
  Opt. B} \textbf{\bibinfo{volume}{4}}, \bibinfo{pages}{272}
  (\bibinfo{year}{2002}).

\bibitem[{\citenamefont{Florescu}(2006)}]{florescu06a}
\bibinfo{author}{\bibfnamefont{L.}~\bibnamefont{Florescu}},
  \bibinfo{journal}{Phys. Rev. A} \textbf{\bibinfo{volume}{74}},
  \bibinfo{pages}{063828} (\bibinfo{year}{2006}).

\bibitem[{\citenamefont{Bienert et~al.}(2007)\citenamefont{Bienert, Torres,
  Zippilli, and Morigi}}]{bienert07a}
\bibinfo{author}{\bibfnamefont{M.}~\bibnamefont{Bienert}},
  \bibinfo{author}{\bibfnamefont{J.~M.} \bibnamefont{Torres}},
  \bibinfo{author}{\bibfnamefont{S.}~\bibnamefont{Zippilli}}, \bibnamefont{and}
  \bibinfo{author}{\bibfnamefont{G.}~\bibnamefont{Morigi}},
  \bibinfo{journal}{Phys. Rev. A} \textbf{\bibinfo{volume}{76}},
  \bibinfo{pages}{013410} (\bibinfo{year}{2007}).

\bibitem[{\citenamefont{Clemens and Rice}(2000)}]{clemens00a}
\bibinfo{author}{\bibfnamefont{J.~P.} \bibnamefont{Clemens}} \bibnamefont{and}
  \bibinfo{author}{\bibfnamefont{P.~R.} \bibnamefont{Rice}},
  \bibinfo{journal}{Phys. Rev. A} \textbf{\bibinfo{volume}{61}},
  \bibinfo{pages}{063810} (\bibinfo{year}{2000}).

\bibitem[{\citenamefont{Perea et~al.}(2004)\citenamefont{Perea, Porras, and
  Tejedor}}]{perea04a}
\bibinfo{author}{\bibfnamefont{J.~I.} \bibnamefont{Perea}},
  \bibinfo{author}{\bibfnamefont{D.}~\bibnamefont{Porras}}, \bibnamefont{and}
  \bibinfo{author}{\bibfnamefont{C.}~\bibnamefont{Tejedor}},
  \bibinfo{journal}{Phys. Rev. B} \textbf{\bibinfo{volume}{70}},
  \bibinfo{pages}{115304} (\bibinfo{year}{2004}).

\bibitem[{\citenamefont{L\"offler et~al.}(1997)\citenamefont{L\"offler, Meyer,
  and Walther}}]{loffler97a}
\bibinfo{author}{\bibfnamefont{M.}~\bibnamefont{L\"offler}},
  \bibinfo{author}{\bibfnamefont{G.~M.} \bibnamefont{Meyer}}, \bibnamefont{and}
  \bibinfo{author}{\bibfnamefont{H.}~\bibnamefont{Walther}},
  \bibinfo{journal}{Phys. Rev. A} \textbf{\bibinfo{volume}{55}},
  \bibinfo{pages}{3923} (\bibinfo{year}{1997}).

\bibitem[{\citenamefont{Clemens et~al.}(2004)\citenamefont{Clemens, Rice, and
  Pedrotti}}]{clemens04a}
\bibinfo{author}{\bibfnamefont{J.~P.} \bibnamefont{Clemens}},
  \bibinfo{author}{\bibfnamefont{P.~R.} \bibnamefont{Rice}}, \bibnamefont{and}
  \bibinfo{author}{\bibfnamefont{L.~M.} \bibnamefont{Pedrotti}},
  \bibinfo{journal}{J. Opt. Soc. Am. B} \textbf{\bibinfo{volume}{21}},
  \bibinfo{pages}{2025} (\bibinfo{year}{2004}).

\bibitem[{\citenamefont{Karlovich and Kilin}(2007)}]{karlovich07a}
\bibinfo{author}{\bibfnamefont{T.~B.} \bibnamefont{Karlovich}}
  \bibnamefont{and} \bibinfo{author}{\bibfnamefont{S.~Y.} \bibnamefont{Kilin}},
  \bibinfo{journal}{Laser Phys.} \textbf{\bibinfo{volume}{103}},
  \bibinfo{pages}{280} (\bibinfo{year}{2007}).

\bibitem[{\citenamefont{Karlovich and Kilin}(2008)}]{karlovich08a}
\bibinfo{author}{\bibfnamefont{T.~B.} \bibnamefont{Karlovich}}
  \bibnamefont{and} \bibinfo{author}{\bibfnamefont{S.~Y.} \bibnamefont{Kilin}},
  \bibinfo{journal}{Laser Phys.} \textbf{\bibinfo{volume}{18}},
  \bibinfo{pages}{783} (\bibinfo{year}{2008}).

\bibitem[{\citenamefont{Hennessy et~al.}(2007)\citenamefont{Hennessy, Badolato,
  Winger, Gerace, Atature, Gulde, {F\u alt}, Hu, and {\u
  Imamo\=glu}}}]{hennessy07a}
\bibinfo{author}{\bibfnamefont{K.}~\bibnamefont{Hennessy}},
  \bibinfo{author}{\bibfnamefont{A.}~\bibnamefont{Badolato}},
  \bibinfo{author}{\bibfnamefont{M.}~\bibnamefont{Winger}},
  \bibinfo{author}{\bibfnamefont{D.}~\bibnamefont{Gerace}},
  \bibinfo{author}{\bibfnamefont{M.}~\bibnamefont{Atature}},
  \bibinfo{author}{\bibfnamefont{S.}~\bibnamefont{Gulde}},
  \bibinfo{author}{\bibfnamefont{S.}~\bibnamefont{{F\u alt}}},
  \bibinfo{author}{\bibfnamefont{E.~L.} \bibnamefont{Hu}}, \bibnamefont{and}
  \bibinfo{author}{\bibfnamefont{A.}~\bibnamefont{{\u Imamo\=glu}}},
  \bibinfo{journal}{Nature} \textbf{\bibinfo{volume}{445}},
  \bibinfo{pages}{896} (\bibinfo{year}{2007}).

\bibitem[{\citenamefont{Feldtmann et~al.}(2006)\citenamefont{Feldtmann,
  Schneebeli, Kira, and Koch}}]{feldtmann06a}
\bibinfo{author}{\bibfnamefont{T.}~\bibnamefont{Feldtmann}},
  \bibinfo{author}{\bibfnamefont{L.}~\bibnamefont{Schneebeli}},
  \bibinfo{author}{\bibfnamefont{M.}~\bibnamefont{Kira}}, \bibnamefont{and}
  \bibinfo{author}{\bibfnamefont{S.~W.} \bibnamefont{Koch}},
  \bibinfo{journal}{Phys. Rev. B} \textbf{\bibinfo{volume}{73}},
  \bibinfo{pages}{155319} (\bibinfo{year}{2006}).

\bibitem[{\citenamefont{Gies et~al.}(2007)\citenamefont{Gies, Wiersig, Lorke,
  and Jahnke}}]{gies07a}
\bibinfo{author}{\bibfnamefont{C.}~\bibnamefont{Gies}},
  \bibinfo{author}{\bibfnamefont{J.}~\bibnamefont{Wiersig}},
  \bibinfo{author}{\bibfnamefont{M.}~\bibnamefont{Lorke}}, \bibnamefont{and}
  \bibinfo{author}{\bibfnamefont{F.}~\bibnamefont{Jahnke}},
  \bibinfo{journal}{Phys. Rev. A} \textbf{\bibinfo{volume}{75}},
  \bibinfo{pages}{013803} (\bibinfo{year}{2007}).

\bibitem[{\citenamefont{Carmichael}(2002)}]{carmichael_book02a}
\bibinfo{author}{\bibfnamefont{H.~J.} \bibnamefont{Carmichael}},
  \emph{\bibinfo{title}{Statistical methods in quantum optics 1}}
  (\bibinfo{publisher}{Springer}, \bibinfo{year}{2002}), \bibinfo{edition}{2nd}
  ed.

\bibitem[{\citenamefont{del Valle et~al.}(2007)\citenamefont{del Valle, Laussy,
  Troiani, and Tejedor}}]{delvalle07b}
\bibinfo{author}{\bibfnamefont{E.}~\bibnamefont{del Valle}},
  \bibinfo{author}{\bibfnamefont{F.~P.} \bibnamefont{Laussy}},
  \bibinfo{author}{\bibfnamefont{F.}~\bibnamefont{Troiani}}, \bibnamefont{and}
  \bibinfo{author}{\bibfnamefont{C.}~\bibnamefont{Tejedor}},
  \bibinfo{journal}{Phys. Rev. B} \textbf{\bibinfo{volume}{76}},
  \bibinfo{pages}{235317} (\bibinfo{year}{2007}).

\bibitem[{\citenamefont{Khitrova et~al.}(2006)\citenamefont{Khitrova, Gibbs,
  Kira, Koch, and Scherer}}]{khitrova06a}
\bibinfo{author}{\bibfnamefont{G.}~\bibnamefont{Khitrova}},
  \bibinfo{author}{\bibfnamefont{H.~M.} \bibnamefont{Gibbs}},
  \bibinfo{author}{\bibfnamefont{M.}~\bibnamefont{Kira}},
  \bibinfo{author}{\bibfnamefont{S.~W.} \bibnamefont{Koch}}, \bibnamefont{and}
  \bibinfo{author}{\bibfnamefont{A.}~\bibnamefont{Scherer}},
  \bibinfo{journal}{Nature Phys.} \textbf{\bibinfo{volume}{2}},
  \bibinfo{pages}{81} (\bibinfo{year}{2006}).

\bibitem[{\citenamefont{Laussy et~al.}(2008{\natexlab{b}})\citenamefont{Laussy,
  del Valle, and Tejedor}}]{laussy08a}
\bibinfo{author}{\bibfnamefont{F.~P.} \bibnamefont{Laussy}},
  \bibinfo{author}{\bibfnamefont{E.}~\bibnamefont{del Valle}},
  \bibnamefont{and} \bibinfo{author}{\bibfnamefont{C.}~\bibnamefont{Tejedor}},
  \bibinfo{journal}{Phys. Rev. Lett.} \textbf{\bibinfo{volume}{101}},
  \bibinfo{pages}{083601} (\bibinfo{year}{2008}{\natexlab{b}}).

\bibitem[{\citenamefont{Benson and Yamamoto}(1999)}]{benson99a}
\bibinfo{author}{\bibfnamefont{O.}~\bibnamefont{Benson}} \bibnamefont{and}
  \bibinfo{author}{\bibfnamefont{Y.}~\bibnamefont{Yamamoto}},
  \bibinfo{journal}{Phys. Rev. A} \textbf{\bibinfo{volume}{59}},
  \bibinfo{pages}{4756} (\bibinfo{year}{1999}).

\bibitem[{\citenamefont{Mu and Savage}(1992)}]{mu92a}
\bibinfo{author}{\bibfnamefont{Y.}~\bibnamefont{Mu}} \bibnamefont{and}
  \bibinfo{author}{\bibfnamefont{C.~M.} \bibnamefont{Savage}},
  \bibinfo{journal}{Phys. Rev. A} \textbf{\bibinfo{volume}{46}},
  \bibinfo{pages}{5944} (\bibinfo{year}{1992}).

\bibitem[{\citenamefont{Ginzel et~al.}(1993)\citenamefont{Ginzel, Briegel,
  Martini, Englert, and Schenzle}}]{gincel93a}
\bibinfo{author}{\bibfnamefont{C.}~\bibnamefont{Ginzel}},
  \bibinfo{author}{\bibfnamefont{H.-J.} \bibnamefont{Briegel}},
  \bibinfo{author}{\bibfnamefont{U.}~\bibnamefont{Martini}},
  \bibinfo{author}{\bibfnamefont{B.-G.} \bibnamefont{Englert}},
  \bibnamefont{and} \bibinfo{author}{\bibfnamefont{A.}~\bibnamefont{Schenzle}},
  \bibinfo{journal}{Phys. Rev. A} \textbf{\bibinfo{volume}{48}},
  \bibinfo{pages}{732} (\bibinfo{year}{1993}).

\bibitem[{\citenamefont{Jones et~al.}(1999)\citenamefont{Jones, Ghose, Clemens,
  Rice, and Pedrotti}}]{jones99a}
\bibinfo{author}{\bibfnamefont{B.}~\bibnamefont{Jones}},
  \bibinfo{author}{\bibfnamefont{S.}~\bibnamefont{Ghose}},
  \bibinfo{author}{\bibfnamefont{J.~P.} \bibnamefont{Clemens}},
  \bibinfo{author}{\bibfnamefont{P.~R.} \bibnamefont{Rice}}, \bibnamefont{and}
  \bibinfo{author}{\bibfnamefont{L.~M.} \bibnamefont{Pedrotti}},
  \bibinfo{journal}{Phys. Rev. A} \textbf{\bibinfo{volume}{60}},
  \bibinfo{pages}{3267} (\bibinfo{year}{1999}).

\bibitem[{\citenamefont{Karlovich and Kilin}(2001)}]{karlovich01a}
\bibinfo{author}{\bibfnamefont{T.~B.} \bibnamefont{Karlovich}}
  \bibnamefont{and} \bibinfo{author}{\bibfnamefont{S.~Y.} \bibnamefont{Kilin}},
  \bibinfo{journal}{Opt. Spectrosc.} \textbf{\bibinfo{volume}{91}},
  \bibinfo{pages}{343} (\bibinfo{year}{2001}).

\bibitem[{\citenamefont{Kozlovskii and Oraevskii}(1999)}]{kozlovskii99a}
\bibinfo{author}{\bibfnamefont{A.}~\bibnamefont{Kozlovskii}} \bibnamefont{and}
  \bibinfo{author}{\bibfnamefont{A.}~\bibnamefont{Oraevskii}},
  \bibinfo{journal}{Sov. Phys. JETP} \textbf{\bibinfo{volume}{88}},
  \bibinfo{pages}{666} (\bibinfo{year}{1999}).

\bibitem[{\citenamefont{Scully and Zubairy}(2002)}]{scully_book02a}
\bibinfo{author}{\bibfnamefont{M.~O.} \bibnamefont{Scully}} \bibnamefont{and}
  \bibinfo{author}{\bibfnamefont{M.~S.} \bibnamefont{Zubairy}},
  \emph{\bibinfo{title}{Quantum optics}} (\bibinfo{publisher}{Cambridge
  University Press}, \bibinfo{year}{2002}).

\end{thebibliography}

\end{document}